\documentclass[journal, twocolumn]{style/IEEEtranTCOM} 

\makeatletter
\def\ps@headings{%
\def\@oddhead{\mbox{}\scriptsize\rightmark \hfil\thepage}%
\def\@evenhead{\scriptsize\thepage \hfil \leftmark\mbox{}}%
\def\@oddfoot{}%
\def\@evenfoot{}}
\makeatother

\makeatletter
\newcommand\resetstackedplots{
\makeatletter
\pgfplots@stacked@isfirstplottrue
\makeatother
\addplot [forget plot,draw=none] coordinates{(150,0) (300,0) (450,0) (600,0) (750,0)};
}
\makeatother

\IEEEoverridecommandlockouts
\usepackage{cuted}
\usepackage{standalone}
\usepackage{threeparttable}
\usepackage[dvipsnames]{xcolor}
\usepackage{calc}
\usepackage[space]{cite}
\usepackage{boxedminipage}
\usepackage{adjustbox}
\usepackage[latin9]{inputenc}
\usepackage{amsfonts}
\usepackage{diagbox}
\usepackage{graphicx}
\usepackage{amssymb}
\usepackage{stmaryrd}
\usepackage{amsmath, amsthm}
\usepackage{autobreak}
\usepackage{verbatim}
\usepackage{makecell}
\usepackage[normalem]{ulem}
\usepackage[noend]{algorithmic}
\usepackage[linesnumbered,ruled,vlined,boxed,commentsnumbered]{algorithm2e}
\usepackage{setspace}

\usepackage{amsmath}
\usepackage{tikz}
\usetikzlibrary{patterns}
\usepackage{pgfplots}
\usepackage{pgfplotstable}
\usepgfplotslibrary{fillbetween}
\usepgfplotslibrary{external} 
\usepackage{graphicx}
\usepackage{subfigure}
\usepackage[abs]{overpic}
\usepackage{multirow}
\usepackage{paralist}
\usepackage{wrapfig}
\usepackage{booktabs}
\usepackage[hyphens]{url}
\usepackage{float}
\usepackage{capt-of}
\usepackage[T1]{fontenc}
\usepackage{siunitx}
\usepackage{mathtools}
\usepackage{array}
\newcolumntype{P}[1]{>{\centering\arraybackslash}p{#1}}

\pgfplotsset{compat=1.3}
\newcommand{\remove}[1]{}

\usepackage{bm}
\usetikzlibrary{patterns}
\definecolor{mygreen}{RGB}{88,212,88}
\definecolor{mypink1}{rgb}{0.858, 0.188, 0.478}
\definecolor{myred}{RGB}{223,42,42}
\definecolor{myyellow1}{RGB}{255,255,204}
\definecolor{myblue}{RGB}{30,144,255}
\definecolor{mygreen0}{RGB}{229,245,224}
\definecolor{mygreen1}{RGB}{161,217,155}
\definecolor{mygreen2}{RGB}{49,163,84}
\definecolor{myblue2}{RGB}{65,182,196}
\definecolor{myblue1}{RGB}{49,130,189}
\definecolor{mypink2}{RGB}{247,104,161}
\definecolor{mypink3}{cmyk}{0, 0.7808, 0.4429, 0.1412}
\definecolor{mygray}{gray}{0.6}
\allowdisplaybreaks 
\theoremstyle{definition}

\usetikzlibrary{patterns}
\usetikzlibrary{shapes.geometric, arrows}

\pgfplotsset{compat=newest}
\usetikzlibrary{positioning}

\tikzstyle{startstop} = [rectangle, rounded corners, minimum width = 1cm, minimum height=0.5cm, text centered, draw = black, fill = red!40]
\tikzstyle{decision} = [diamond, aspect = 3,text centered,draw=black,fill = green!30 ]
\tikzstyle{process} = [rectangle, minimum width=1.5cm, minimum height=0.5cm, text centered, draw=black, fill = yellow!50]
\tikzstyle{arrow} = [->,>=stealth]

\usepackage[pscoord]{eso-pic}

\begin{document}

\title{Throughput Maximization Leveraging Just-Enough SNR Margin and Channel Spacing Optimization}
 

\author{Cao~Chen,
        Fen~Zhou,~\IEEEmembership{Senior~Member,~IEEE}, Yuanhao Liu, and
         Shilin~Xiao

\thanks{The work is jointly supported by Eiffel Scholarship (No. 895145D), China Scholarship Council (No. 201806230093 and No. 202006960046), National Nature Science Fund of China (No.61775137, No.62071295, No.61431009, and No.61433009) and the National "863" Hi-tech Project of China.}
\thanks{Cao Chen is with the State Key Laboratory of Advanced Optical Communication Systems and Networks, Shanghai Jiao Tong University, Shanghai, 200240, China. He is also with the  CERI-LIA in Avignon University, France (email: cao.chen@alumni.univ-avignon.fr).}

\thanks{Fen Zhou is with IMT Lille Douai, Institut Mines-T\'el\'ecom, University of Lille, Center for Digital Systems, F-59000 Lille, France 
(email: fen.zhou@imt-lille-douai.fr).
}


\thanks{Yuanhao Liu is with the State Key Laboratory of Integrated Service Networks, School of Telecommunications Engineering, Xidian University, Xi'an, 710071, China.  He is also with the  CERI-LIA in Avignon University, France (email: yuanhao\_liu@foxmail.com).}

\thanks{Shilin Xiao is with the State Key Laboratory of Advanced Optical Communication Systems and Networks, Shanghai Jiao Tong University, Shanghai, 200240, China (email: slxiao@sjtu.edu.cn).}

}

\markboth{Journal of \LaTeX\ Class Files,~Vol.~14, No.~8, August~20XX}%
{Cao CHEN \MakeLowercase{\textit{et al.}}: Bare Demo of IEEEtran.cls for IEEE Journals}
%


\maketitle

\begin{abstract}

Flexible optical network is a promising technology to accommodate high-capacity demands in next-generation networks. To ensure uninterrupted communication, existing lightpath provisioning schemes are mainly done with the assumption of worst-case resource under-provisioning  and fixed channel spacing, which preserves an excessive signal-to-noise ratio (SNR) margin. However, under a resource over-provisioning scenario, the excessive SNR margin restricts the transmission bit-rate \textcolor{black}{or transmission reach}, leading to physical layer resource waste and stranded transmission capacity. To tackle this challenging problem, we leverage an iterative feedback tuning algorithm to provide a just-enough SNR margin, so as to maximize the network throughput. 
Specifically, the proposed algorithm is implemented in three steps. First, starting from the high SNR margin setup, we establish an integer linear programming model as well as a heuristic algorithm to maximize the network throughput by solving the problem of routing, modulation format, forward error correction, baud-rate selection, and spectrum assignment. Second, we optimize the channel spacing of the lightpaths obtained from the previous step, thereby increasing the available physical layer resources. Finally, we iteratively reduce the SNR margin of each lightpath until the network throughput cannot be increased. 
Through numerical simulations, we confirm the throughput improvement in different networks and with different baud-rates. In particular, we find that our algorithm enables over 20\% relative gain when network resource is over-provisioned, compared to the traditional method preserving an excessive SNR margin.
 
\end{abstract}

\section{Introduction}
\IEEEPARstart{A}{ccording} to recent traffic reports, network traffic (fueled by network services like video on demand, file sharing, online gaming, video conferencing, etc.) are still growing exponentially in today's Internet\cite{cisco2017}.
\textcolor{black}{
The constant traffic growth relies on optical networks. In optical networks, the resource allocation consists of finding a proper lightpath and assigning adequate spectrum resources. Traditionally, each channel in wavelength-division multiplexing (WDM) networks has to follow a rigid fixed-sized grid, e.g. 50~GHz. Recently, in the new paradigm of flexible optical networks (FONs), each channel can choose multiple contiguous frequency slots (FSs) with the size of 12.5~GHz, and freely adjust the center frequency. More recently, with the advance of coherent detection technology and commercially available digital-to-analog converters (DACs), a large number of transceiver's parameters become possible to further improve the spectral efficiency, including baud-rate, modulation format (MF), forward error correction (FEC) overhead, probabilistic shaping, etc\cite{SGSC20,Bosc19}. The development of optical networks enables us to reduce the spectrum resource waste, increase spectral efficiency, and obtain the higher transmission capacity\cite{DaLS15,IvBS15}. However, it is not easy to jointly optimize these parameters.
}
 
\textcolor{black}{
In addition, the selection criteria of these flexible parameters must satisfy the quality of transmission (QoT) requirement. Each lightpath experiences not only the amplified spontaneous emission (ASE) noise of optical amplifiers, but also the fiber nonlinear interference  (NLI) from other lightpaths that share a common link. To avoid the complex calculation of NLI while guaranteeing uninterrupted communication, the conventional practice mostly assumes that each lightpath experiences the most NLI under resource underprovisioning scenario, which preserves an excessive signal-to-noise (SNR) margin. The high SNR margin or the overestimation of NLI potentially inhibits the capacity or restricts the maximal transmission reach. To this end, NLI-aware resource allocation techniques have been proposed to reduce the overestimation to exploit the stranded capacity\cite{BeSa13, XYAP19, YADW17, WBMN19, ZhWA15}. This operation mode is also named as \textit{low margin network} in \cite{Yvan17,SaVI19, Ciena19}, or \textit{just-enough SNR margin} in this paper. }

\textcolor{black}{Nevertheless, baud-rate was assumed to ideally match the occupied spectrum resource in most prior studies\cite{BeSa13, XYAP19, YADW17, WBMN19, ZhWA15}, thus they only need to consider the route, modulation format, and spectrum assignment (RMSA). Limited by the frequency grid, different baud-rates (or the multi-baud-rates) can be accommodated by the same channel once the allocated spectrum resource is sufficient\cite{Infinera19}. Recent studies have also shown that the same baud-rate using different bandwidths will experience baud-rate-related impairments, including the filter narrowing effect\cite{MRZG15, CMLB17}, or the impairments caused by the resolution of DACs\cite{GSSE20}. Thus, the resource allocation incorporating the baud-rate selection maps into a new problem of routing, modulation format, FEC, baud-rate, and spectrum assignment, which was also investigated by \cite{RBMT17,STCT19,SMCF15, PCSP19}. Although some researchers have proposed the efficient heuristic algorithm optimizing these parameters under the low SNR margin mode\cite{SaCV18, SCQP17, Pedr17}, there is no integer linear programming (ILP) model on the joint optimization of the route, MF, FEC, baud-rate, and spectrum assignment, especially aiming at improving the transmission capacity. This problem could become more complex when accounting for the various impairments.
}

\textcolor{black}{
Optimizing the physical layer parameter is also important to mitigate the NLI and improve network capacity\cite{SaVI19, Auge13, PBCK13}. As reported by the recent studies of the physical layer (e.g., Gaussian Noise model on the Nyquist WDM channel \cite{Pogg12,JoAg14}), the NLI between channels relates to the baud-rate, power, and different spectrum positions (channel spacing and channel order). There is a large volume of published studies on the power optimization \cite{YADW17, IvBS15, RoKB16, BAKK19} and the channel order optimization (the different relative orders of channels) \cite{YAWP15, BAKK19}. The last factor---channel spacing---is generally implemented by setting a guard band, or more simply, by setting a fixed channel spacing. Throughout this paper, we use the terms ``guard band'' and ``channel spacing'' interchangeably. By setting a proper channel spacing, we can obtain the desired metrics, such as improving the SNR performance\cite{MRZG15,RRBZ20} or ensuring a proper security level\cite{KFZW16}. Thus, it can be formulated as another optimization problem, \textit{channel spacing optimization problem}, as we will study in this paper. Although the transmission performance of different channel spacing has been widely studied \cite{BCCP11} in the physical layer, a few studies concentrate on the optimization from the network operator's view. While the study in \cite{NaTM13} has reported the best channel spacing strategy aiming at reducing the cost, it is still unclear the transmission bit-rate gain of channel spacing optimization. }

\textcolor{black}{Traditionally, to obtain a higher transmission capacity, one may take the evaluation metric, \textit{minimizing the maximum spectral usage}, to spare spectrum resources for other lightpaths. However, restricting the spectrum usage potentially encourages narrow channel spacing and neglects the benefit of large channel spacing on mitigating the filter narrowing effect or NLI. Thus, the proposed channel spacing optimization strategy allows the channel to be freely adjusted among the available spectrum resources without guard band constraints or spectrum resource usage constraints, to earn the ultimate physical layer performance improvement. Recently, the method of individually optimizing the center frequency seems to be a powerful technique \cite{YAWP15,YADW17} to optimize the channel spacing rather than using the heuristic algorithm\cite{RRBZ20}. However, this novel method may face difficulty when increasing the number of lightpaths. To lower the complexity, we further propose a linear programming (LP) model based on the nearest neighbor channel, successfully extending the application case to the general mesh optical network with a large number of lightpaths. 
}

\begin{figure}[!tbp]
\centering
\includegraphics[width=0.5\textwidth]{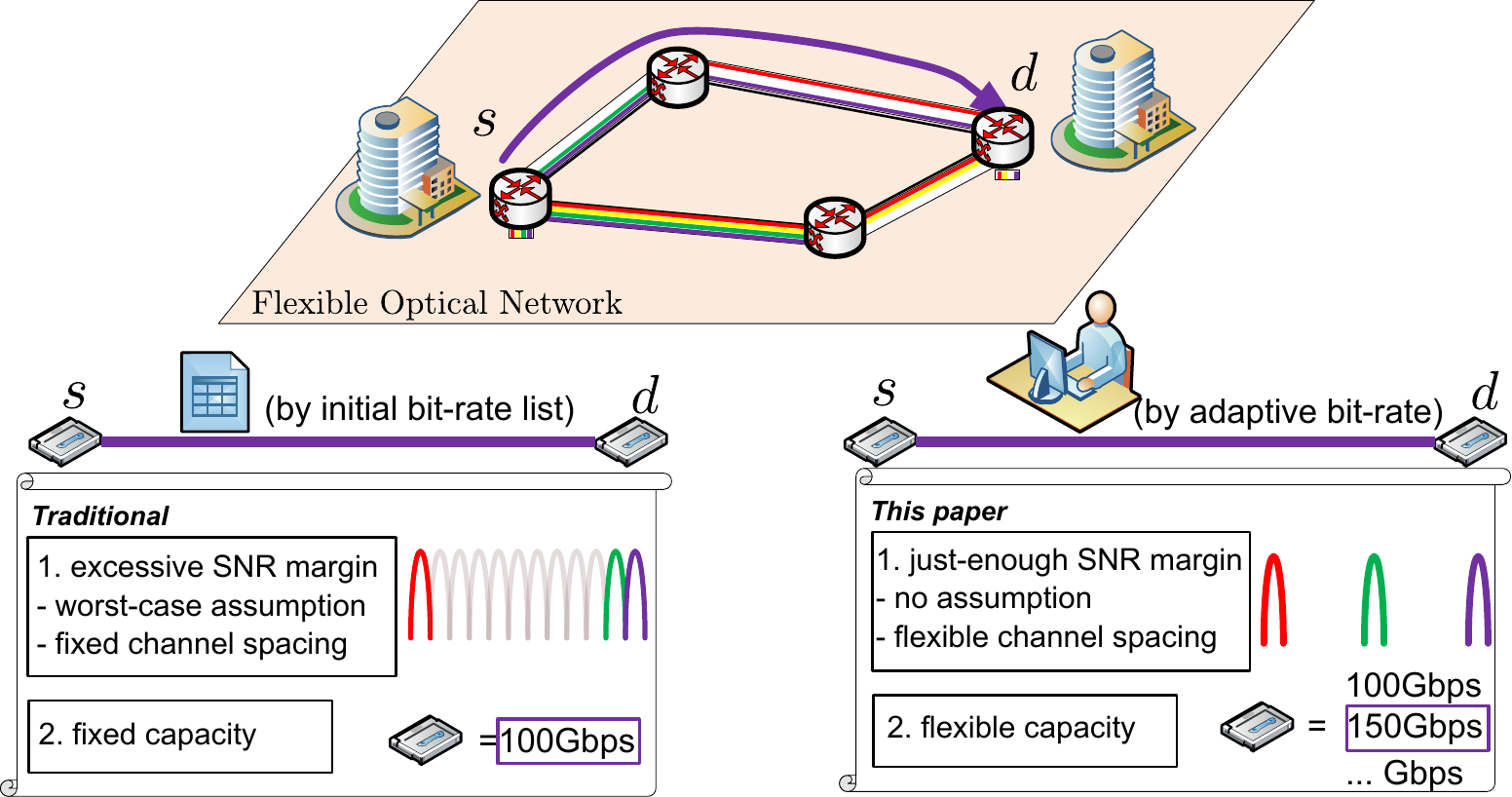}

\caption{Illustration of two operation modes in FON.}
\label{fig: concept}
\end{figure}

\textcolor{black}{In this paper}, we aim at maximizing the network throughput in a static FON by leveraging just-enough SNR margin and channel spacing optimization. \textcolor{black}{As shown in Fig.~\ref{fig: concept}, only the lowest transmission capacity can be adopted by the lightpath connecting $(s,d)$ when considering the excessive SNR margin and fixed channel spacing, namely 100~Gbps in the traditional operation mode. The transmission capacity can increase up to 150 Gbps or larger if leveraging the just-enough SNR margin provisioning and channel spacing optimization. To implement this function in large networks,} we propose an iterative feedback tuning algorithm to provide a just-enough SNR margin for each lightpath. Specifically, our algorithm is implemented in \textit{three} steps. First, starting from a high SNR margin, we establish an  ILP model and a heuristic algorithm that maximizes the network throughput. Unlike the existing lightpath provisioning that \textcolor{black}{addresses the RMSA, the baud-rate is also jointly optimized}. Second, \textcolor{black}{we provide a low-complexity channel spacing optimization model to mitigate the NLI, which is implemented by using the nearest neighbor channel}. This model \textcolor{black}{also} extends the application case from a ring network into a mesh network compared to our previous study\cite{ONDM2020}. Finally, we iteratively reduce the SNR margin of each lightpath to the just-enough level to increase the  throughput. The main contributions are summarized as follows,

\begin{itemize}

\item We propose to \textit{maximize the throughput by leveraging just-enough SNR margin and channel spacing optimization}. 
 \textcolor{black}{ Through analyzing and evaluating the impact of different lightpath's parameters (route, MF, FEC, baud-rate, spectrum assignment), we determine the value of just-enough SNR margin accordingly. As far as we know, this is the first time that the flexible transceiver's parameters are jointly optimized while considering the NLI.}
\item \textcolor{black}{
We introduce an LP model based on the nearest neighbor channel to optimize the channel spacing, resulting in a low-complexity method that applies to a large number of lightpaths in both ring and mesh optical networks. Besides, we compare the strategy with the widely used channel spacing strategies (fixed and candidate channel spacing) in terms of the SNR performance and achievable throughput.
}
    \item Next, for a given resource over-provisioning scenario, we devise an iterative feedback tuning algorithm that can efficiently solve the \textit{throughput maximization leveraging just-enough SNR margin and channel spacing optimization problem}. Specifically,  we reduce the SNR margin by adjusting a slack \textcolor{black}{parameter} iteratively so that we can obtain the just-enough SNR margin.
    \item Finally, through extensive simulations in different network topologies and with different baud-rates, we confirm the throughput improvement with the help of the just-enough SNR margin and channel spacing optimization. Specifically, we observe that the relative throughput gain ratio (no smaller than 20\% \textcolor{black}{in our case}) is more evident in resource over-provisioning scenarios, while the absolute throughput gain is more evident when network traffic load is medium.
\end{itemize}

The rest of this paper is organized as follows. In Sec. \ref{sec: problem}, we present the preliminary  information of the lightpath provisioning in FONs and the \textit{throughput maximization leveraging just-enough SNR margin and the channel spacing optimization problem}. Next, we propose an iterative feedback tuning algorithm to solve this problem in Sec. \ref{sec: iterative}. To reduce the complexity of the lightpath provisioning, we also present an efficient heuristic in Sec. \ref{sec: heursic_maxthroughput}. Illustrative numerical results are presented in Sec. \ref{sec: simulation}. Finally, Sec. \ref{sec: conclusions} concludes this paper.

\section{Throughput Maximization Leveraging Just-enough SNR Margin and Channel Spacing Optimization Problem }\label{sec: problem}

In this section, we present the studied \textit{throughput maximization leveraging just-enough SNR margin and channel spacing optimization problem}. First, we provide the necessary information about the lightpath provisioning of FONs and explain how we calculate the \textit{SNR}. Then, we describe the throughput maximization problem and present a benchmark method that utilizes the fixed baud-rate. Next, we present the key problem to be solved in this paper. Finally, a case study is used to illustrate the throughput improvement with the help of channel spacing optimization and just-enough SNR margin.

\subsection{Network model}

We use $G=(V,E)$ to denote an FON, where $V$ and $E$ are the node and link set, respectively.  A link $l\in E$ represents a fiber $uv$ ($u, v\in V$). On that fiber, the available spectrum resource is $F$ (unit: GHz), while the number of available FSs is $W$ ($F$=$f_{grid}\cdot W$, $f_{grid}$ is 12.5~GHz bandwidth). A bit-mask $b[1...W]$ of size $W$ is introduced to represent the occupied FS status of a fiber link, where $b[w]$=1 means the $w^\text{th}$ FS is occupied. We also use $b_*[w]$ to represent whether the entity `*' uses the $w^\text{th}$ FS, where `*' could be either a lightpath $p$, or a link $l$, or a channel $ch$.
 
We assume that each lightpath $p$ connecting a node pair $(s,d)$ can adopt a transceiver $t$ operating at a baud-rate $R_t$, which needs to occupy a certain number of FSs denoted by $B_t$ (refer to \cite[Table 3]{OIF2019}). \textcolor{black}{The adopted transceiver is assumed with the rectangular Nyquist spectrum on a  channel $ch$.}
\textcolor{black}{Note that the different baud-rates $R_t$ could use the same amount of FS(s) $B_t$, or the same baud-rate could use the different amount of FS(s) only if $B_t\cdot f_{grid} \geq R_t$\cite{Infinera19}, we determine the available bit-rate by the baud-rate $R_t$ rather than occupied $B_t$.}
Also, the bit-rate depends on the spectral efficiency of adopted MF and FEC (an example table that shows the SNR threshold and bit-rate of transceiver at 32 Gbaud can refer to  \cite[Table I]{IWLS16}). For ease of expression, we introduce the term \textit{transmission mode} to refer to a combination of MF and FEC. Such a transmission mode can be adopted by the lightpath if the SNR is no smaller than the SNR threshold of joint MF and FEC, i.e. QoT is satisfied. Next, we describe how we calculate the SNR of each lightpath.

For each lightpath, the SNR will be \textcolor{black}{degraded by the ASE noise}  and NLI. Supposing the transceiver of lightpath $p$ uses a launch power spectral density (PSD) $G_p$, its SNR is stated as follows\cite{JoAg14,IWLS16},
\textcolor{black}{
\begin{subequations}
\begin{align}
    \label{eq: OSNR expression}
    SNR_{p} &= \frac{R_t \cdot G_p}{f_{grid}\cdot B_t\cdot (G_p^{ASE} + G^{NLI}_p)} \\
    &= \frac{R_t \cdot G_p}{f_{grid}\cdot B_t\cdot G_p^{ASE}} \cdot \frac{1}{1+\frac{ G^{NLI}_p}{G_p^{ASE}}}\\
    \label{eq: SNRbest}
    &= SNR_{p}^{best} \cdot \frac{1}{1+\frac{  G^{NLI}_p}{G_p^{ASE}}}  = SNR_{p}^{best} \frac{1}{\mathcal{X}_p}
    \end{align}
\end{subequations}
}
where $G_p^{ASE}$ is the accumulated PSD of ASE noise in optical amplifiers, $G_p^{NLI}$ is the accumulated PSD of NLI on fibers,  $SNR^{best}_p=\frac{R_t \cdot G_p}{f_{grid}\cdot B_t\cdot G_p^{ASE}}$ is the received \textcolor{black}{best-case} SNR without NLI, \textcolor{black}{$\mathcal{X}_{p} = 1+\frac{G_p^{NLI}}{G_p^{ASE}}$ is the relative strength of NLI.}
\textcolor{black}{By assuming a coherent receiver with adaptive digital signal processors and a matched receiver filter, the noise equivalent bandwidth of $f_{grid}\cdot B_t$ is equal to the baud-rate $R_t$. Thus, we can ignore the impact of bandwidth and baud-rate on the SNR evaluation in the following.}
We also assume a uniform launch PSD for all transceivers. 

\textcolor{black}{
The NLI strength $\mathcal{X}_{p}$ in Eq.~(\ref{eq: SNRbest}) depends on the occupied channels of lightpath. 
}For ease of discussion, we present the log form of 
SNR as follows, 
\textcolor{black}{
\begin{align}
\label{eq: log_SNR_best}
SNR_{p,\text{[dB]}} =  SNR_{p,\text{[dB]}}^{best} -  \mathcal{X}_{p,\text{[dB]}}  
\end{align}
where $\mathcal{X}_{p,\text{[dB]}}=10\log_{\text{10}}\left(1+\frac{G_p^{NLI}}{G_p^{ASE}}\right)$.
}
Next, we explain the calculation of ASE noise and NLI. 
\begin{itemize}
    \item \textbf{ASE}. We use the following equation to  calculate the ASE noise of optical amplifiers,
\begin{align}  
G_p^{ASE} &= N_{p} \left( 10^{\alpha   L_{span}}-1\right) n_{sp}   h  \nu
\end{align}
where $N_{p}$ is the number of spans, $L_{span}$ (km) is the length of one span between two optical amplifiers, $\alpha$ is the fiber attenuation factor, $n_{sp}$ is the noise figure of optical amplifiers, $h$ is the Planck constant, and $\nu$ is the absolute frequency of optical signal at 1,550~nm.

\item \textbf{NLI}. We adopt the dilog method in  \cite[Eq.~(11)]{JoAg14} \textcolor{black}{that calculates} the NLI \textcolor{black}{based on channel},
\begin{align}
  \label{eq: NLIspan}
  G^{NLI}_{p} &= N_{p}   \eta_{ch}^{SCI}  {G_p}^3 +  G_p  \sum_{ch'} \eta_{ch,ch'}^{XCI} (f_{ch,ch'})    N_{p,p'}   {G_{p'}}^2
\end{align}
where $ch$ and $ch'$ are the occupied \textcolor{black}{channel} of lightpath $p$ and $p'$ respectively, $N_{p,p'}$ is the number of common spans of lightpath $p$ and $p'$, $\eta^{SCI}_{ch}$ and  $\eta^{XCI}_{ch,ch'}$ are the self-channel interference (SCI) efficiency and  cross-channel interference (XCI)  efficiency  respectively,   $f_{ch,ch'}$ is the \textit{channel spacing}  of two lightpaths (this paper uses the center frequency difference, i.e., $f_{ch,ch'}=|f_{ch}-f_{ch'}|$). An example curve of the XCI efficiency at  different channel spacing can refer to \cite[Fig.~1]{IvBS14}.
Meanwhile, we ignore the impact of SCI, as it can be compensated by digital back propagation  \cite{IpKa08} \textcolor{black}{ and has a weak impact on the parameter of channel spacing}\cite{JoAg14}.
 
\end{itemize}

\subsection{Throughput maximization problem}\label{sec: sub_throughput}

Let us now look at the \textcolor{black}{network throughput}. We first define a traffic demand matrix $\bm{D}$, \textcolor{black}{where each item $D_{s,d}$ denotes the bit-rate demand between each node pair $(s,d)$}.  We also define another network provisioning capacity matrix $\bm{T}$, where each item $T_{s,d}$ is \textcolor{black}{the transmission capacity of the lightpaths between node pair $(s,d)$.} To ensure that the traffic demand could be fully accommodated, \textcolor{black}{the  bit-rate demand should be no bigger than the transmission capacity, i.e. $D_{s,d}\leq T_{s,d}$}.  \textcolor{black}{The network throughput $TH$ in this paper is the sum of all bit-rate demands, i.e. $TH = \sum_{(s,d)}D_{s,d}$}.

\textcolor{black}{
As this paper considers a static lightpath provisioning, we begin by assuming a constant traffic distribution, namely the normalized traffic demand matrix $\hat{\bm{D}}$ is constant,  where $\hat{D}_{s,d}=\frac{D_{s,d}}{\sum_{(s,d)} D_{s,d}}$. Thus, we have $TH\cdot \hat{D}_{s,d} = D_{s,d} \leq T_{s,d}$.}

Given a normalized traffic demand matrix \textcolor{black}{$\bm{\hat{D}}$} and a transmission mode set $\mathcal{M}$, the objective function is expressed as follows,  
\begin{align}
\label{eq: obj_throughput_max}
\bm{\max} \left\{\sum_{(s,d)} D_{s,d} \right\}
\end{align}

In this problem, we need to individually design the number of lightpaths to implement the network provisioning capacity $T_{s,d}$ for each node pair. \textcolor{black}{For each lightpath, the route, MF, FEC, baud-rate, and spectrum assignment needs to satisfy the following constraints.}

\begin{itemize}
\item\textbf{C1 - QoT constraint}: Each lightpath should guarantee a satisfied QoT. \textcolor{black}{The QoT constraint can be expressed either by the QoT metric $\mathcal{Q}_p=\frac{SNR^{}_{p}}{SNR^{threshold}_{m}}\geq 1$, or by its log form using  Eq.~(\ref{eq: log_SNR_best}),}
\textcolor{black}{
\begin{align}
\label{eq: prob_con1}
SNR^{best}_{p,\text{[dB]}} -  \mathcal{X}_{p,\text{[dB]}} -  SNR_{m,\text{[dB]}}^{threshold}  \geq 0,
\end{align}
}
where $SNR_m^{threshold}$ is the SNR threshold of the adopted transmission mode $m$. 
\textcolor{black}{ The $SNR_m^{threshold}$ of a transmission could vary with the  baud-rates\cite{GSSE20} or the aging factor. Here, we assume ideal case that it only varies with the adopted transmission mode. Recall that the second term $\mathcal{X}_p$ in Eq.~(\ref{eq: prob_con1}) depends on the route and spectrum positioning of other undetermined lightpaths, which could dramatically increase the difficulty of lightpath provisioning.}
A simple way to eliminate \textcolor{black}{$\mathcal{X}_p$ in Eq.~(\ref{eq: prob_con1}) is using a high SNR margin $M_p$}. Such a margin needs to compensate all possible NLI of lightpath $p$. Thus, we give the alternative constraint, \textit{SNR margin requirement}. 

\item\textbf{C1* - SNR margin requirement}:
\textcolor{black}{
\begin{align}
\label{eq: prob_con1_2}
SNR^{best}_{p,\text{[dB]}}  -  M_{p,\text{[dB]}} -  SNR_{m,\text{[dB]}}^{threshold} \geq 0, 
\end{align}
}
where $M_p$ is the \textit{SNR margin} used for compensating the unknown NLI. Generally, the  SNR margin \textcolor{black}{$M_p\geq \mathcal{X}_p^{worst}$, 
where $\mathcal{X}_p^{worst}$ is calculated under the worst-case assumption in which other spectrum resources of the links along lightpath are fully occupied by different channels.} With the SNR margin requirement constraint, we can neglect the complex physical layer calculation in QoT constraint, \textcolor{black}{and subsequently focus on the lightpath provisioning by using a conservative transmission reach or the conservative spectral efficiency.} 
\item\textbf{C2 - one baud-rate and one transmission mode constraint}: Each lightpath can adopt one and only one baud-rate and one transmission mode.

\item\textbf{C3 - spectrum contiguity constraint}: 
For each used baud-rate $R_t$, we need to allocate $B_t$ contiguous FSs, namely $b_{ch}[1...W]=[\underbrace{0\cdots 0\overbrace{11\cdots 11}^{B_t}0\cdots 0}_W]$. \textcolor{black}{It needs to ensure the size of spectrum allocated is sufficient, that is $f_{grid}\cdot B_t \geq R_t$}.

\item\textbf{C4 -  spectrum continuity constraint}: \textcolor{black}{The occupied FSs of the channel for a lightpath should be identical} on each link.
\item\textbf{C5 -  spectrum non-overlapping constraint}: The occupied FSs of two lightpaths should not overlap. Alternatively, each FS of a link can be used at most once.
\end{itemize}

Let us discuss the constraints (\textbf{C1}) and (\textbf{C1*}). 
Traditional method takes constraint (\textbf{C1*}) rather than (\textbf{C1}). The benefit is that lightpath provisioning can be greatly simplified by splitting the physical layer calculation and lightpath provisioning \textcolor{black}{when} the SNR margin is enough for compensating the NLI. The limitation is that the improper SNR margin $M_p$ requires each lightpath to adopt the low order transmission mode with \textcolor{black}{a} low SNR threshold. As a result, the transmission bit-rate of each lightpath is strictly constrained, especially under the resource over-provisioning scenario where the actual NLI may be far less than the worst-case NLI. 

Here, to make it clear, we present a benchmark method that preserves excessive SNR margin\cite{IvBS15}. It mainly includes two parts, a lightpath precalculation based on constraint (\textbf{C1*}) and an ILP model based on constraints (\textbf{C2})-(\textbf{C5}). \textcolor{black}{Using the pre-calculated lightpaths, the ILP model calculates the maximal throughput by jointly solving the routing, transmission mode, and spectrum assignment problem (on a fixed wavelength grid). Additionally, we  use another constraint to define the mentioned scenario of resource over-provisioning, where the actual NLI is much lower than the preserved SNR margin. This constraint limits the current maximum FS index $W_{cur}$, $W_{cur}\leq W$.} 

\subsubsection{Lightpath precalculation}

This part \textcolor{black}{provides the candidate lightpaths to be used in the following optimization model. These candidate lightpaths are obtained by using the $K$-shortest path algorithm\cite{Yen71} to generate the different routes, using the channel index method \cite{VKRC12} to generate the possible channel locations, and using $M_p = \mathcal{X}_p^{worst}$ to generate the available transmission modes. For a certain lightpath, its transmission capacity will be zero if the Eq.~(\ref{eq: prob_con1_2}) is not satisfied.}

\subsubsection{Maximize  throughput}\label{sec: subsub_max}
This part maximizes the network throughput based on (\textbf{C2})-(\textbf{C5}).

\noindent\textbf{Parameters}
\begin{itemize}
\item $\hat{D}_{s,d}$, normalized traffic demand ratio of node pair $(s,d)$.
\item  $ch\in CH$, channel index of different contiguous FSs. For example, for the transceiver using two contiguous FSs, $CH  =\{[110\cdots], [0110\cdots], \cdots, [\cdots 011]\}$.
\item \textcolor{black}{$b_{ch}[w]$, equals 1 if the channel $ch$ uses the $w^\text{th}$ FS, 0 otherwise.}
\item $p_{s,d,k}^{ch,m} \in \mathcal{P}$, lightpath ID of the $k^\text{th}$  route of $(s,d)$ using channel $ch$ and transmission mode $m$.
\item $C_{s,d,k}^{ch,m}$, transmission bit-rate of the  lightpath $p_{s,d,k}^{ch,m}$ if it satisfies Eq.~(\ref{eq: prob_con1_2}), 0 otherwise.
\item $\beta_{s,d,k,l}$, equals 1 if the $k^\text{th}$ route of node pair $(s,d)$ uses link $l$, 0 otherwise.
\end{itemize}
\textbf{Variables}
\begin{itemize} 
\item \textcolor{black}{$D_{s,d}$, bit-rate demand of node pair $(s,d)$.}
\item \textcolor{black}{$T_{s,d}$, provisioning capacity of node pair $(s,d)$.}
\item $\delta_{s,d,k}^{ch,m}$, equals 1 if the lightpath $p_{s,d,k}^{ch,m}$ is adopted, 0 otherwise.
\item $TH$, network throughput.
\end{itemize}
\textbf{Objective}
\begin{subequations}
\label{model: P1_SB}
\begin{align}
 & \textcolor{black}{\bm{\max}\limits_{\delta_{s,d,k}^{ch,m},TH}} \quad  TH
 \quad \boldsymbol{\texttt{(Max TH)}} && \nonumber  \\
 \label{eq: con1_inthroughput_normalized}
\textbf{\textit{s.t.~~~}} &  D_{s,d}=TH\cdot \hat{D}_{s,d}, \quad\quad \forall s, d \neq s &&\\
\label{eq: conAddi_inthroughput_DT}
&  T_{s,d} = \sum_{k,ch,m} \delta_{s,d,k}^{ch,m} \cdot C_{s,d,k}^{ch,m}, \quad \forall s, d \neq s  &&\\
\label{eq: con1_inthroughput_b}
 &  \textcolor{black}{D_{s,d} \leq  T_{s,d}, \quad \quad \quad \quad \forall s, d \neq s}&&\\
\label{eq: con3_inthroughput_spectrumSlot_b}
& \sum_{s,d\neq s,k,ch,m} \delta_{s,d,k}^{ch,m} \cdot \beta_{s,d,k,l} \cdot b_{ch}[w] \leq 1, \quad \forall l,w && \\
\label{eq: con4_inthroughput_Fcur_b}
& \delta_{s,d,k}^{ch,m} \cdot b_{ch}[w] =0. \quad \forall s,d,k,ch, W_{cur} < w \leq W,m && 
\end{align}
\end{subequations}
The objective is to maximize the network throughput. 
Constraints (\ref{eq: con1_inthroughput_normalized}) ensure that the bit-rate demand follows the distribution of the given normalized traffic demand matrix $\hat{\bm{D}}$ for each node pair. \textcolor{black}{
Constraints (\ref{eq: conAddi_inthroughput_DT}) obtain the transmission capacity between $(s,d)$.}  Constraints (\ref{eq: con1_inthroughput_b}) ensure that the traffic demand could be fully accommodated by the network provisioning capacity. Constraints (\ref{eq: con3_inthroughput_spectrumSlot_b}) ensure that each FS of a link can be used at most once, which also \textcolor{black}{guarantee only one transmission mode for a lightpath}. Constraints (\ref{eq: con4_inthroughput_Fcur_b}) are used to simulate the resource over-provisioning scenario by the FS index $W_{cur}$. Finally, we see that the constraints (\textbf{C3}) and (\textbf{C4}) are ignored due to the usage of pre-calculated lightpaths.
   
\subsection{Throughput maximization leveraging just-enough SNR margin provisioning and channel spacing optimization}

\textcolor{black}{
The benchmark only allows optimization in networks configured by a single baud-rate. This function may be insufficient for the application in future optical networks that allows the coexistence of flexible baud-rate optical networks\cite{PeCP20}. Thus,} the first question is that i) \textit{how to efficiently use the flexible baud-rate transceiver for lightpath provisioning}? 
In addition,  the benchmark method may inevitably generate redundant lightpaths. \textcolor{black}{For example, for some node pair $(s,d)$ with a short distance, their lightpaths have a high provisioning capacity, easily satisfies the bit-rate demand, and probably leaves idle spectrum resources on the connected links.
It is inevitable to generate additional lightpaths that still connect the same node pair using these idle spectrum resources.} \textcolor{black}{Nevertheless, the local capacity improvement of several node pairs may not improve the global scaling factor of $TH$.} Meanwhile, these lightpaths will however introduce extra network cost and NLI. Thus, the second question is that ii) \textit{how to remove these  lightpaths while guaranteeing the \textcolor{black}{current} maximal network throughput}?

\textcolor{black}{ Returning to the discussion of Eqs.~(\ref{eq: prob_con1}) and (\ref{eq: prob_con1_2}), we can now say that one important approach to increase the network throughput is to allow the high-order transmission modes for each lightpath, i.e., increasing more lightpaths $p_{s,d,k}^{ch,m}$ that pass the SNR margin requirement in precalculation process. Mitigating the NLI strength of $\mathcal{X}_p$ in Eq.~(\ref{eq: prob_con1}) or lowering the SNR margin $M_p$ in Eq.~(\ref{eq: prob_con1_2}) both provide the possibility.  On the one hand, to mitigate NLI, prior studies have optimized the power and channel order allocation\cite{YADW17,RBGR19,BAKK19}. This paper will study it from another perspective, how to use the idle spectrum resources in resource overpromising scenario, namely optimizing the channel spacing. Thus, the third question is that \textit{iii) how to minimize the NLI to allow each lightpath to adopt a higher-order transmission mode?} On the other hand, lowering the SNR performance has been an important subject of the prior studies\cite{IvBS15,SCQP17,YADW17,WBMN19}. It seems that the low SNR margin provisioning is incorporated with the lightpath provisioning simultaneously, which could incur difficulty in lightpath provisioning. In addition, to the best of our knowledge, the resource allocation with the flexible transceiver parameters has not been employed under the resource overprovisioning scenario\cite{IvBS15}. Thus, the fourth question is that \textit{iv) how to find a properly preserved SNR margin to suit  the resource overprovisioning scenario to allow the lightpath to adopt a higher-order transmission mode?}
}

Question i) is solved by incorporating the baud-rate into the lightpath pre-calculation. Question ii) is solved by \textcolor{black}{adding a posterior  optimization similar to \texttt{Max TH} but using a modified objective function that minimizes the total lightpaths.} The exact technique for these two questions will be presented in Sec.~\ref{sec: iterative}. \textcolor{black}{
The last two questions, iii) and iv) both involve the physical layer optimization. Here, we present the two problems sequentially.
}

\subsubsection{Channel spacing optimization}  Given a lightpath set $\mathcal{P}_{adopt}$ with determined route, determined baud-rate, and determined transmission mode, we want to enhance the SNR by minimizing the NLI for each lightpath. \textcolor{black}{When minimizing the NLI, the QoT metric $\mathcal{Q}_p$ of different lightpaths is considered. Thus, we leverage the three terms on the left side of Eq.~(\ref{eq: prob_con1}) as the objective. The objective function is expressed as follows,}
\textcolor{black}{
\begin{align}
  \nonumber  & \bm{\min}_{f_{ch}} \left\{\max_{p\in \mathcal{P}_{adopt}} \left\{ \mathcal{X}_{p,\text{[dB]}} +  SNR_{m,\text{[dB]}}^{threshold} - SNR^{best}_{p, \text{[dB]}}   \right\}\right\}\\
\textbf{\textit{s.t.~~~}} & \textbf{(C1), (C5)}  \nonumber 
\end{align}
} 

\subsubsection{Just-enough SNR margin provisioning}\label{sec: JP} \textcolor{black}{Given a normalized traffic demand matrix, a transmission mode set $\mathcal{M}$,  and a baud-rate set $\mathcal{T}$, the objective function is expressed as follows,
}
\begin{subequations}
\label{model: Just-enough}
\begin{align}
& \bm{\max}\limits_{\delta_{s,d,k}^{ch,m,R},f_{ch},M_p} \quad  TH \quad && \nonumber  \\
\textbf{\textit{s.t.~~~}} & \textbf{(C1)},\textbf{(C2)}-\textbf{(C5)} \nonumber 
\end{align}
\end{subequations}

\textcolor{black}{
Unlike the benchmark method, the just-enough SNR margin provisioning needs to determine a proper SNR margin $M_p$ for each lightpath. Besides, the route $(s,d,k)$, channel $ch$, transmission mode $m$, and baud-rate $R$, as well as a proper channel location $f_{ch}$ for each lightpath are to be optimized.
}

\subsubsection{A small instance}

\begin{figure}[!htbp]
    \centering
\if@twocolumn%
        \includegraphics[]{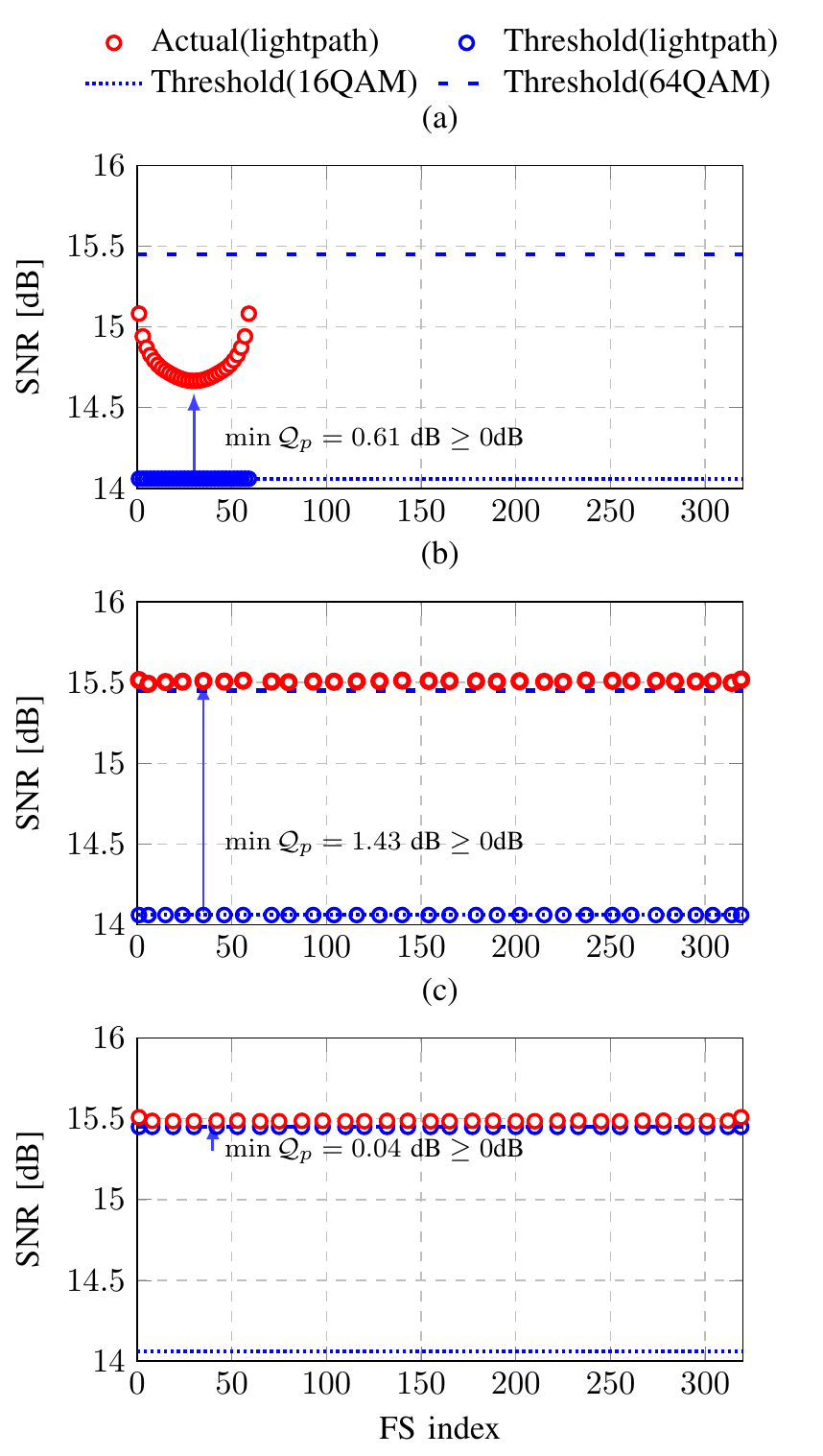}
\else
\fi
 
    \caption{An example of lightpath provisioning \textit{leveraging just-enough SNR margin and channel spacing optimization} in a point-to-point network with 600~km ($W_{cur}$=60~FSs, $W$=320~FSs, PSD=\textcolor{black}{15}~$\mu$W/GHz in this example). The lightpaths need to satisfy the QoT, i.e. Eq.~(\ref{eq: prob_con1}) holds. (a) Traditional  provisioning  through preserving  an  excessive  SNR  margin (throughput is 30$\times$112.5~Gbps); (b) Provisioning with channel spacing optimization; (c) Provisioning with just-enough SNR margin (throughput is 30$\times$125~Gbps). }
    \label{fig: demo}
\end{figure}

\textcolor{black}{We provide a small instance to illustrate the throughput improvement when adopting the just-enough SNR margin provisioning and channel spacing optimization. This instance assumes 30 lightpaths in a point-to-point network with available spectrum resources of 4,000~GHz ($F$=4,000~GHz, $W$=320~FSs).}
In Fig.~\ref{fig: demo}, we plot the SNR distribution of 30 lightpaths provisioned by using three different techniques: \textcolor{black}{(a)} traditional provisioning with an excessive SNR margin, \textcolor{black}{(b)} provisioning with channel spacing optimization, and \textcolor{black}{(c)} provisioning with just-enough SNR margin. Each red circle denotes a lightpath that uses two contiguous FSs \textcolor{black}{at 16~Gbaud}, where $x$-axis is the spectrum position and $y$-axis is the lightpath's actual SNR. They can adopt either the transmission mode (16QAM, FEC~=~0.92) or (64QAM, FEC~=~0.68)\cite{IWLS16}.  \textcolor{black}{The  bit-rates are set with 112.5 and 125~Gbps to match the 16~Gbaud transceiver in our assumption\cite{IWLS16}.} \textcolor{black}{ More details of the physical layer parameters (fiber type, EDFA noise figure, and span length) can refer to the simulation setup in Sec.~\ref{sec: simulation_setup}. The uniform PSD $G_{opt}$ (15.03$\mu$W/GHz) is given by using a strategy similar to LOGON in \cite[Eq.~(6)]{PBCC13} among 4,000~GHz, in which we assume other spectrum resources are fully occupied by all possible channels.
}

In Fig.~\ref{fig: demo}(a), 30 lightpaths all adopt 16QAM and provide the bit-rate of 112.5~Gbps. The QoT is satisfied because all SNRs are over the SNR threshold \textcolor{black}{($\min \mathcal{Q}_p =0.61$~dB > $0$~dB)}.  Next, if we apply the channel spacing optimization, as shown in Fig.~\ref{fig: demo}(b), the SNRs of these
lightpaths rise because of the less NLI \textcolor{black}{($\min  \mathcal{Q}_p =1.43$~dB)}. In Fig.~\ref{fig: demo}(c), by using a lower SNR margin, we can upgrade the transmission mode to 64QAM which provides a larger bit-rate of 125 Gbps while guaranteeing the QoT for all lightpaths \textcolor{black}{($\min \mathcal{Q}_p =0.04$~dB)}. Therefore, we can increase the network throughput from
30$\times$112.5~Gbps  in Fig.~\ref{fig: demo}(a) to 30$\times$125~Gbps in Fig.~\ref{fig: demo}(c).

\textcolor{black}{It should be also mentioned that the maximum SNR margin improvement of the traditional fixed channel spacing is 0.02~dB ($\min \mathcal{Q}_p$=0.02~dB) in Fig.~\ref{fig: demo}(c) (not shown in the figure for the sake of clarity), which is achieved by setting 125~GHz frequency grid.  
This metric is still less than the 0.04~dB of flexible channel spacing.}

\textcolor{black}{Through the example in Fig.~\ref{fig: demo}}, we can upgrade the transmission mode and increase the bit-rate by properly setting the channel spacing and carefully adjusting the SNR margin. However, it \textcolor{black}{seems} not easy to \textcolor{black}{simultaneously adjust the channel spacing and SNR  margin}, especially for a large number of lightpaths. To tackle this challenge, we design an iterative feedback tuning algorithm to \textcolor{black}{iteratively} adjust these parameters until the just-enough SNR margin is found.

\section{
Iterative Feedback Tuning Algorithm and  Mathematical Models}\label{sec: iterative}

\begin{figure}[!htbp]
\centering
 \includegraphics[width=8.5cm]{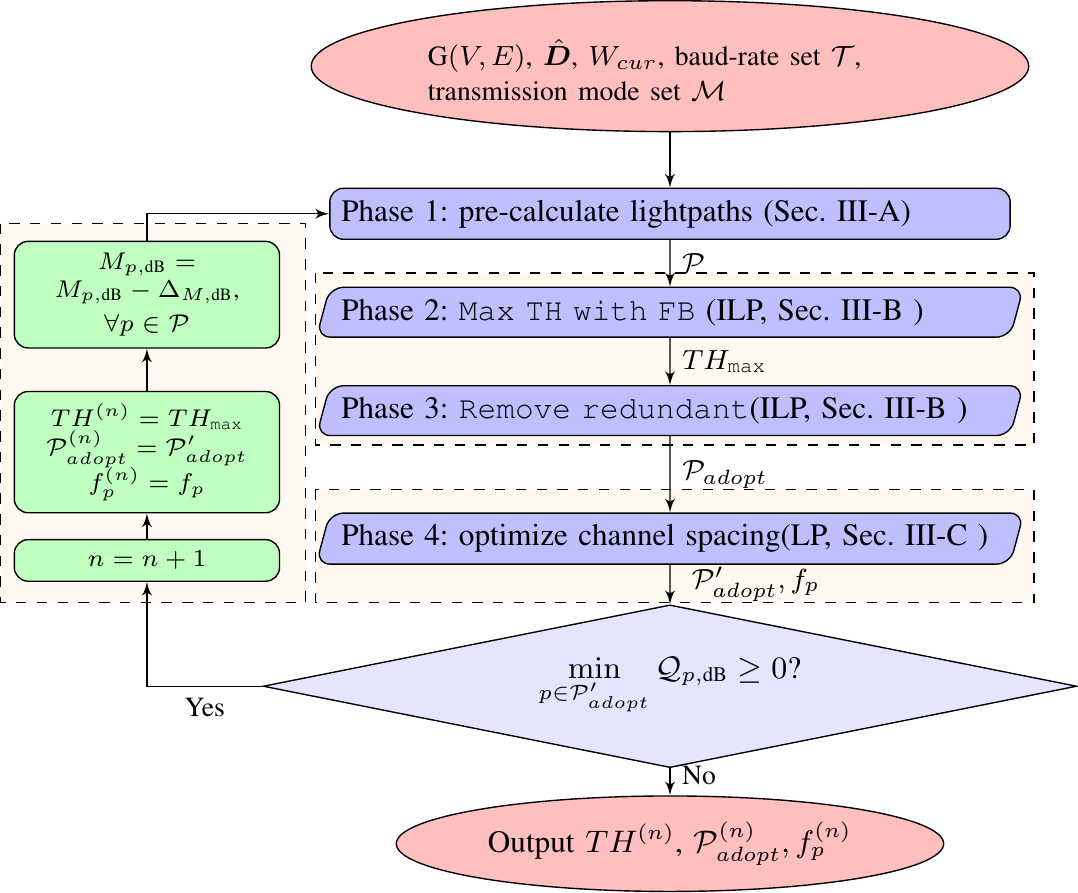}
    \caption{Flowchart of our iterative feedback tuning algorithm.}
    \label{fig: flowchart}
\end{figure}

In this section,  we present the algorithm that solves the \textit{throughput maximization leveraging just-enough SNR margin and channel spacing optimization problem}. The flowchart is illustrated in Fig.~\ref{fig: flowchart}.  The first two phases mainly follow the benchmark method in Sec. \ref{sec: sub_throughput}. The difference \textcolor{black}{lies in} that the flexible baud-rates are incorporated. Then, we use  \texttt{Phase 3} to remove \textcolor{black}{the} redundant lightpaths of \texttt{Phase 2}. Next, in \texttt{Phase 4}, we optimize the channel spacing 
through an LP model. 
Motivated by \cite{IvBS15}, we try to determine the just-enough SNR margin $M_p$ with a unit gradient ratio step of $\Delta_M$.  
Compared to last iteration, \textcolor{black}{more high-order transmission modes are permitted for the precalculated lightpaths in \texttt{Phase 1}, thus increase the possibility of \texttt{Phase 2} using high-order transmission mode, and consequently obtain  higher network throughput. Finally, the algorithm terminates once there exists  a lightpath in the network that violates the QoT constraint.}

\subsection{Lightpath precalculation with flexible baud-rate}

\begin{itemize}
    \item \textbf{Input}: topology $G(V,E)$,   maximal FS index $W_{cur}$, transceiver set $\mathcal{T}$, transmission mode set $\mathcal{M}$, SNR margin $M_p$ (initialized by \textcolor{black}{$\mathcal{X}_p^{worst}$}).
    \item \textbf{Output}: lightpath set \textcolor{black}{$p_{s,d,k}^{ch,m,R}\in\mathcal{P}$}, capacity set \textcolor{black}{$\mathcal{C}_{s,d,k}^{ch,m,R}\in\mathcal{C}$},  boolean indicator $\beta_{s,d,k,l}$.
    \item \textbf{Process}: it stores the lightpath $p$ that satisfies the SNR margin requirement in Eq.~(\ref{eq: prob_con1_2}), where  $M_p$ acts as a slack variable for each iteration. \textcolor{black}{The satisfied lightpaths } are stored into the candidate lightpath set $\mathcal{P}$ and the capacity set $\mathcal{C}$. Different from the lightpath precalculation in Sec. \ref{sec: sub_throughput}, here, we use \textcolor{black}{$p_{s,d,k}^{ch,m,R}$} to denote the lightpath ID rather than $p_{s,d,k}^{ch,m}$. The channel index set $CH$ is also extended to \textcolor{black}{accommodate the different baud-rates}.  
\end{itemize}

\subsection{Throughput maximization with flexible baud-rate}\label{sec: p2 and p3}

\subsubsection{Phase 2} 

\begin{itemize}
    \item \textbf{Input}: candidate lightpath set $\mathcal{P}$, capacity set $\mathcal{C}$, boolean indicator $\beta_{s,d,k,l}$, normalized traffic demand matrix $\bm{\hat{D}}$.
    \item \textbf{Output}: network throughput $TH_{\texttt{max}}$.
    \item \textbf{Process}: this phase calculates the maximal network throughput through the following model,
    
\end{itemize}
\begin{subequations}
\label{model: P1_FB}
\begin{align}
 & \textcolor{black}{\bm{\max}\limits_{\delta_{s,d,k}^{ch,m,R},TH}} \quad  TH \quad \boldsymbol{\texttt{(Max TH with FB)}} && \nonumber  \\
 \label{eq: con1_inthroughput}
 \textbf{\textit{s.t.~~~}} &  D_{s,d}=TH\cdot \hat{D}_{s,d}, \quad\quad \forall s, d \neq s &&\\
  &  T_{s,d} = \sum_{k,ch,m,R} \textcolor{black}{\delta_{s,d,k}^{ch,m,R}} \cdot \textcolor{black}{C_{s,d,k}^{ch,m,R}}, \quad \forall s, d \neq s  &&\\
 &  \textcolor{black}{D_{s,d} \leq  T_{s,d}, \quad\quad\quad\quad \forall s, d \neq s  } &&\\
\label{eq: con3_inthroughput_spectrumSlot}
& \textcolor{black}{\sum_{s,d\neq s,k,ch,m,R} \delta_{s,d,k}^{ch,m,R}} \cdot \beta_{s,d,k,l} \cdot b_{ch}[w] \leq 1, \quad \forall l,w && \\
\label{eq: con4_inthroughput_Fcur}
& \textcolor{black}{\delta_{s,d,k}^{ch,m,R}} \cdot b_{ch}[w] =0. \quad \forall s,d,k,ch,\textcolor{black}{R}, W_{cur} < w \leq W,m && 
\end{align}
\end{subequations}

The description of these constraints can refer to the benchmark method in Sec.~\ref{sec: sub_throughput}. Different from that, we have replaced the variable $\delta_{s,d,k}^{ch,m}$ with \textcolor{black}{$\delta_{s,d,k}^{ch,m,R}$} in order to optimize the baud-rate. \textcolor{black}{Thus, the route $(s,d,k)$, spectrum position $ch$, transmission mode $m$, and baud-rate $R$ in \texttt{Max TH with FB} are jointly optimized.}

\subsubsection{Phase 3} 
\begin{itemize}
    \item \textbf{Input}: candidate lightpath set $\mathcal{P}$, capacity set $\mathcal{C}$, boolean indicator $\beta_{s,d,k,l}$, normalized traffic demand matrix $\bm{\hat{D}}$,  network throughput $TH_{\texttt{max}}$.
    \item \textbf{Output}: adopted lightpath set $\mathcal{P}_{adopt}$.
    \item \textbf{Process}: this phase removes the redundant lightpaths of previous model. Based on the model \texttt{Max TH with FB}, \textcolor{black}{we use the objective of minimizing the total number of lightpaths while maintaining the network throughput $TH_{\max}$.}
\end{itemize}
\begin{subequations}
\label{model: Reducing_lightpaths}
\begin{align}
&\textcolor{black}{\bm{\min}\limits_{\delta_{s,d,k}^{ch,m,R}} \sum_{s,d\neq s,k,ch,m,R} \delta_{s,d,k}^{ch,m,R} \quad \boldsymbol{\texttt{(Remove Redundant)}}} \nonumber\\
\textbf{\textit{s.t.~~~}} & (\ref{eq: con1_inthroughput}) - (\ref{eq: con4_inthroughput_Fcur}) \nonumber \\
\label{eq: con2_inreducing_Tmax}
 & TH \geq TH_{\mathtt{max}}
\end{align} 

\end{subequations}

The objective is to minimize the total number of lightpaths in the network.  Constraint (\ref{eq: con2_inreducing_Tmax}) maintains  the throughput $TH_{\texttt{max}}$ that is obtained from previous model.  \textcolor{black}{The redundant lightpaths can be removed through this model. 
}

\subsection{Channel spacing optimization}
\begin{itemize}
    \item \textbf{Input}: adopted lightpath set $\mathcal{P}_{adopt}$.
    \item \textbf{Output}: optimal center frequency $f_p$, \textcolor{black}{$\mathcal{P}'_{adopt}$ on the optimal center frequency $f_p$}. 
    \item \textbf{Process}: \textcolor{black}{$\bm{\min}\limits_{f_p}  \left\{\max\limits_{p\in \mathcal{P}_{adopt}} ( \mathcal{X}_p + SNR_{m,\text{[dB]}}^{threshold}  - SNR^{best}_{p,\text{[dB]}}
    ) \right\}$. }
\end{itemize}

\textcolor{black}{
The objective function can be also expressed in the linear form $\bm{\min}\limits_{f_p} \left\{\max \left\{ \frac{SNR^{threshold}_m}{SNR_p^{best}} \cdot  \mathcal{X}_p \right\}\right\}$.
} \textcolor{black}{We take the center frequency as a continuous variable among the available spectrum resources $[0,F]$.} Thus, the center frequency of each lightpath can be individually optimized by an LP model. Meanwhile, we have to resolve the issue that the relationship between XCI efficiency and channel spacing in Eq.~(\ref{eq: NLIspan}) is nonlinear. To this end, we use a piecewise-linear fitting function\cite{YADW17}. Finally, the continuous variable $f_p$ is rounded to fit the frequency grid of $f_{grid}$ (unit:~GHz). \textcolor{black}{The frequency mapping process in the final step is decomposed into two-step to accelerate the optimization.} 

The linear approximation function for fitting the XCI efficiency is expressed as follows\cite{YADW17}, \begin{align}
\label{eq: XCI_fit}
 \bar{\eta}_{B_{p},B_{p'}}^{XCI} (f_{ch,ch'}) = \max_{1\leq q \leq Q} \left( a_q^{B_{p},B_{p'}} f_{ch,ch'}   + b_q^{B_{p},B_{p'}} \right)
\end{align}
where $Q$ is the number of fitting segments ($Q$~=~5), $a_q^{B_{p},B_{p'}}$ and $b_q^{B_{p},B_{p'}}$ are the coefficients calculated by the algorithm in \cite{MaBo09} \textcolor{black}{based on the XCI efficiency of the channels from $p'$ to $p$ with frequency step of 1~GHz in Eq.~(\ref{eq: NLIspan}).}

\noindent\textbf{Parameters}
\begin{itemize}
    
    \item \textcolor{black}{$f_{grid}$, bandwidth of an FS (12.5~GHz).} 
    \item $x_{p,l}$, equals 1 if \textcolor{black}{lightpath} $p$ uses link $l$, 0 otherwise.
    \item $y_{p,p'}$, equals 1 if lightpath $p$ and $p'$ share a common fiber link, 0 otherwise. 
    \textcolor{black}{\item $\gamma_p$, spectrum position of a lightpath $p$. Here, we take the starting FS index. }
    
    \item $u_{p,p'}$, \textcolor{black}{equals 1 if $\gamma_p$ is bigger than $\gamma_{p'}$}, 0 otherwise.
    \item $B_p$, number of occupied FSs of a lightpath $p$.
    \textcolor{black}{\item $m_p$, transmission mode of $p$.}
    \textcolor{black}{\item $a_q^{B_{p},B_{p'}}, b_q^{B_{p},B_{p'}}$, coefficients of piecewise linear fitting function.}
    \item  $G_{span}^{ASE}$, PSD of ASE between two optical amplifiers, i.e.  per one span.
\end{itemize}

\noindent\textbf{Variables}
\begin{itemize}
    \item $f_p$, center frequency of lightpath $p$ that is relative to the optical signal frequency $\nu$.
    \item $\bar{\eta}_{p,p'}^{XCI}$, linear approximation of the XCI efficiency from lightpath $p'$ to $p$.
    \item $\mathcal{X}_{p,l}$, the strength of NLI of lightpath $p$ on link $l$.
    \item $\mathcal{X}_{}^{network}$, \textcolor{black}{the maximum NLI strength in a network considering the QoT metric.}
    \end{itemize}
\textbf{Objective function}
\begin{subequations}
\label{eq: model_frequency_optimization}
\begin{align}
 &  \textcolor{black}{\bm{\min}\limits_{f_{p},\bar{\eta}_{p,p'}^{XCI}}} \quad  \mathcal{X}_{}^{network} \quad \boldsymbol{\texttt{(Phase 4)}} && \nonumber\\
\label{eq: con1_frequency_mapping_margin}
\textbf{\textit{s.t.~~~}} & \textcolor{black}{\frac{{SNR_{m_p}^{threshold}}}{SNR_p^{best}}} \cdot  \sum_l  \mathcal{X}_{p,l} \leq  \mathcal{X}_{}^{network}, \quad \forall p &&  \\
\label{eq: con2_frequench_mapping_pli}
& x_{p,l} \left\{ 1 + \frac{ \sum_{p'\neq p} x_{p',l} \cdot {G_p}^3 \cdot \bar{\eta}_{p,p'}^{XCI}}{G_{span}^{ASE}}  \right\}  \leq \mathcal{X}_{p,l},  \forall p,l && \\
\label{eq: con3_frequench_mapping_pli}
\begin{split}
y_{p,p'} &\left\{ \begin{aligned}
& \left[ a_q^{B_p,B_{p'}} \left( f_p-f_{p'}\right) + b_q^{B_p,B_{p'}} \right] \\
	& \quad \quad \cdot u_{p,p'} -\bar{\eta}_{p,{p'}}^{XCI}
  \end{aligned} \right\} \leq 0,  \\ & \quad\quad\quad\quad\quad\quad\quad\quad\quad\quad\quad\quad\quad\quad \forall p,p'\neq p, q 
\end{split}
  \\
\label{eq: con4_frequench_mapping_pli}
\begin{split}
y_{p,p'} &\left\{ \begin{aligned}
	& \left[ a_q^{B_p,B_{p'}}  \left( f_{p'} - f_p\right) + b_q^{B_p,B_{p'}} \right] \\
	& \quad\quad \cdot (1-u_{p,p'}) -\bar{\eta}_{p,{p'}}^{XCI}
  \end{aligned} \right\} \leq 0, \quad \\
  & \quad\quad\quad\quad\quad\quad\quad\quad\quad\quad\quad\quad\quad\quad \forall p, p'\neq p, q
\end{split}\\ 
\label{eq: con5_frequench_mapping_non_overlapping}
\begin{split}
0\leq&\left[\left( \frac{f_p}{f_{grid}}-\frac{B_p}{2}\right) - \left( \frac{f_{p'}}{f_{grid}} + \frac{B_{p'}}{2}\right)\right] y_{p,p'}  u_{p,p'} \\ 
& \quad\quad\quad\quad\quad\quad\quad\quad\quad\quad\quad\quad\quad\quad \forall p, p'\neq p  
\end{split}
\\
\label{eq: con6_boundary_bonud}
 & \frac{f_{p}}{f_{grid}} + \frac{B_p}{2} \leq W, 0\leq \frac{f_{p}}{f_{grid}} - \frac{B_p}{2}. \quad \forall p&&
 \end{align}

\end{subequations}

 The objective is to minimize the maximum NLI strength for all lightpaths in the network. Such a ratio, denoted by $\mathcal{X}^{network}$, is calculated by constraints (\ref{eq: con1_frequency_mapping_margin}). Constraints (\ref{eq: con2_frequench_mapping_pli})  calculate the NLI strength for the lightpath $p$  on link $l$. Constraints (\ref{eq: con3_frequench_mapping_pli}) and (\ref{eq: con4_frequench_mapping_pli}) calculate the XCI efficiency $\bar{\eta}_{p, p'}^{XCI}$ between lightpath $p$ and $p'$ \textcolor{black}{using $|f_p-f_{p'}|$}. Constraints (\ref{eq: con5_frequench_mapping_non_overlapping}) guarantee that the occupied spectrum resources of two lightpaths are non-overlapping if they use a common link. Constraints (\ref{eq: con6_boundary_bonud}) restrict the spectrum resources among the spectrum interval $[0,F]$. \textcolor{black}{
 Note that the QoT constraint of (\textbf{C1}) can be dealt with independently. Thus, we place this  constraint as a check function in the final step of our algorithm (see Sec.~\ref{sec: tuning SNR margin}).
 Here, the main variables to be optimized in \texttt{Phase 4} are the center frequencies $f_p$ and $\bar{\eta}_{p,p'}^{XCI}$.
 } 
 
 \textcolor{black}{
 One advantage of the above model is to obtain the ideal maximal performance improvement of individually optimizing the channel spacing. However, there are certain drawbacks associated with the method of \textit{fully estimating the NLI}. For instance, assuming a simple case in a point-to-point link, the number of constraints in Eqs.~(\ref{eq: con3_frequench_mapping_pli}) and (\ref{eq: con4_frequench_mapping_pli}) could approach to $\mathcal{O}(Q\cdot|\mathcal{P}_{adopt}|^2)$. Such a square relationship probably restricts the optimization to a small number of lightpaths, which could be much difficult when facing a large number of lightpaths.}

\textcolor{black}{
To reduce the issue of complexity,  we propose a \textit{nearest neighborhood channel} estimation method. This estimation method needs to use two parameters that describe the nearest channel relationship between lightpaths, $y^{(1)}_{p,p'}$ and $y^{(2)}_{p,p'}$. $y^{(1)}_{p,p'}$ equals 1 if the lightpath $p'$ (shares the same link of $p$, i.e., $y_{p,p'}=1$) has the minimum starting FS difference with lightpath $p$ on either side (left and right sides are separately employed), 0 otherwise. $y^{(2)}_{p,p'}$ equals 1 if there exists a third lightpath $p''$ that satisfies three conditions, i) $y^{(1)}_{p,p''}=1$, ii) $y^{(1)}_{p'',p'}=1$, and iii) $y_{p,p'}=1$. Finally, with the obtained parameter of $y^{(1)}_{p,p'} + y^{(2)}_{p,p'}$, we replace the parameter $y_{p,p'}$ in Eqs.~(\ref{eq: con3_frequench_mapping_pli}) and (\ref{eq: con4_frequench_mapping_pli}). By doing so, we just need to consider the NLI of the nearest four channels rather than all lightpaths share a common link, thus lowering the complexity from $\mathcal{O}(Q\cdot|\mathcal{P}_{adopt}|^2)$ to $\mathcal{O}(Q\cdot|\mathcal{P}_{adopt}|\cdot 4)$ for the point-to-point case. The effectiveness of this method has been experimentally shown in the literature \cite[Fig.~4]{BCCP11}, which shows that the impact of XCI efficiency could be negligible when the channel spacing is sufficiently large (over two bandwidths).
} 

\textcolor{black}{
Currently, the widely used channel spacing strategies are still the fixed or candidate channel spacing strategies. To make a comparison with these strategies, we provide the following constraints into  \texttt{Phase 4} to implement the function, respectively. Here, we refer to our optimization strategy as CSO, fixed channel spacing strategy as FIX, and candidate channel spacing strategy as CAN. In addition, CAN(opt) differs from CAN(random) in whether to use the objective function of minimizing the NLI. Also, FIX adopts no objective function. For both FIX and CAN, the channel spacing set is denoted as $\mathcal{H}$.
\begin{itemize}
    \item \textbf{FIX}: Each channel follows a fixed channel spacing $h$, $h\in \mathcal{H} (|\mathcal{H}|=1)$. For the networks that support heterogeneous bandwidths, we allow the small bandwidth channel to follow integer times of the channel spacing $h$, while the large bandwidth channel follows the given channel spacing.  Thus, for any $p\in \mathcal{P}_{adopt}, p'\in \mathcal{P}_{adopt}/\{p\}$,
    \begin{align}
        0&\leq (f_{p}-f_{p'} - h) u_{p,p'} \cdot y_{p,p'}^{(1)}\leq (n_{p,p'}-1)\cdot h.\\
        1&\leq n_{p,p'} \leq |\mathcal{T}_B|, \quad\quad n_{p,p'}\in\mathbb{N}^{+}.
    \end{align}
    Here, $|\mathcal{T}_B|$ is the number of heterogeneous bandwidths $B_t$ in the transceiver set $\mathcal{T}$.
\end{itemize}
\begin{itemize}
    \item \textbf{CAN(opt)}: We allow the channel spacing $f_p-f_{p'}$ to vary among the interval $[\inf(\mathcal{H}), \sup(\mathcal{H})]$ if the channel of $p'$ is the nearest neighborhood channel of $p$. Thus, for any $p\in \mathcal{P}_{adopt}, p'\in \mathcal{P}_{adopt}/\{p\}$,
\end{itemize}
\begin{align}
    0\leq (f_p - f_{p'} - \inf{(\mathcal{H})}) u_{p,p'} \cdot y_{p,p'}^{(1)}\leq \sup(\mathcal{H})- \inf{(\mathcal{H})}
\end{align}
\begin{itemize}
    \item \textbf{CAN(random)}:  Similar to CAN(opt), we allow the channel spacing $f_p-f_{p'}$ to vary among the set $[h_p, \sup(\mathcal{H})]$ if the channel of $p'$ is the nearest neigbor channel of $p$, where $h_p\in \mathcal{H}$ is a random value. 
    Thus, for any $p\in \mathcal{P}_{adopt}, p'\in \mathcal{P}_{adopt}/\{p\}$, 
\begin{align}
    0 \leq (f_p - f_{p'} - h_p) u_{p,p'} \cdot y_{p,p'}^{(1)}\leq \sup(\mathcal{H})- h_p
\end{align}
\end{itemize}
}

\textcolor{black}{
\subsection{Tuning SNR margin}\label{sec: tuning SNR margin}}

\textcolor{black}{
Finally, we deal with the excessive SNR margin. After the lightpath selection (\texttt{Phase 2} and \texttt{Phase 3}) and the channel spacing optimization (\texttt{Phase 4}), the actual SNR performance of the lightpath in $\mathcal{P}'_{adopt}$ could be higher than its worst-case assumption. The enhanced SNR performance allows the lightpath to choose higher-order transmission mode (see Fig.~\ref{fig: demo}(b) and Fig.~\ref{fig: demo}(c)). To allow the lightpath to adopt a higher-order transmission mode, we can achieve it by either of the options. Option i), upgrade the transmission mode of current lightpath set $\mathcal{P}_{adopt}'$, or option ii), lower the SNR margin in the \texttt{Phase 1}. This paper takes the latter one to globally optimize the candidate lightpath set $\mathcal{P}$. To this end, we create a control loop connecting the \texttt{Phase 1} and store the current feasible  results.} 

\textcolor{black}{We describe the detailed process of tuning the SNR margin $M_p$. At each iteration, $M_p$ decreases with a unit step $\Delta M_p$ (0.5~dB in this paper). As the lowered SNR margin $M_p$ employs in the pre-calculation process, all lightpaths are promising to  adopt a higher-order transmission mode, thus increasing the average spectral efficiency of each lightpath in \texttt{Phase 2} and obtaining a higher network throughput. Through the iterations, the QoT metric $\min \mathcal{Q}_p$ after \texttt{Phase 4} also decreases. Once there exists a lightpath that violates the QoT constraint ($\exists p\in \mathcal{P}'_{adopt}$, $\mathcal{Q}_{p,\text{dB}}<0$), we output the last stored results.  
}

\section{Heuristic for Throughput Maximization} \label{sec: heursic_maxthroughput}

In this section, we devise a sequential loading algorithm to efficiently solve the throughput maximization problem in Sec. \ref{sec: sub_throughput}. The baud-rate optimization is also incorporated.

According to the common sense, we can increase the provisioning capacity between $(s,d)$ by using the following options,  i) upgrade the transmission mode of the working lightpath, ii) adjust the baud-rate or the route of the working lightpath, iii) establish a new lightpath directly. We denote by $\Delta_{TH}^{mod}$, $\Delta_{TH}^{trans}$, and $\Delta_{TH}^{new}$ the bit-rate gain of these three options, respectively. \textcolor{black}{As indicated by the adopted Gaussian Noise \cite{JoAg14} model}, option i) increases the bit-rate without incurring additional NLI to other lightpaths, 
which is regarded as the first choice.  Options ii) and iii) both \textcolor{black}{degrade} the \textcolor{black}{QoT} of other existing lightpaths. To reduce the extra interference from the new lightpath, we give priority to option ii) and then choose option iii).

\begin{algorithm}[!htbp]
\SetKwInOut{Input}{Input}
\SetKwInOut{Output}{Output}
\SetKw{KwAnd}{and}
\SetKw{KwSuch}{s.t.}
\SetKwRepeat{Do}{do}{while}
\Input{\textcolor{black}{$G(V,E), W_{cur}$, $W$, $\hat{D}_{s,d}$, lightpath set $\mathcal{P}$, capacity set $\mathcal{C}$, boolean indicator $\beta_{s,d,k,l}$}}
\Output{$TH$, $\mathcal{P}_{adopt}$}

Set the initial network throughput $TH$ and step $\Delta_{TH}$, \label{alg: step}
Terminate = false\;
\While{!Terminate}{
\For{Node pair $(s,d)$} {
\label{alg: start}
Calculate current \textcolor{black}{provisioning capacity} $T_{s,d}$\;
Option i): upgrade transmission mode $m$  for existing lightpaths to obtain $\Delta^{mod}_{TH}$\;\label{alg: option 1}
\If{$T_{s,d} + \Delta^{mod}_{TH} < (TH+\Delta_{TH})\cdot \textcolor{black}{\hat{D}_{s,d}}$\label{alg: if1}}
{
Try either one of the following: 
\\Option ii): adjust route $k$,  baud-rate \textcolor{black}{$R$}, or move the channel $ch$ to gain $\Delta^{trans}_{TH}$ 
\;\label{alg: option 2}
Option iii): establish a new lightpath to gain $\Delta^{new}_{TH}$\; \label{alg: option 3}

\If{$T_{s,d}+\Delta^{mod}_{TH} + \max (\Delta^{new}_{TH}, \Delta^{trans}_{TH}) <(TH+\Delta_{TH})\cdot  \textcolor{black}{\hat{D}_{s,d}}$}{
Go to line~\ref{alg: output_network_throughput} to terminate the algorithm\;\label{alg: terminate}
}
\Else{
Upgrade the lightpaths of options ii) and iii) into lightpath set $\mathcal{P}_{adopt}$\;\label{alg: end}
}

}

}
$TH = TH+\Delta_{TH}$\;

}
Output $TH$ and the lightpath set $\mathcal{P}_{adopt}$\label{alg: output_network_throughput}\;

\caption{Sequential loading algorithm for throughput maximization}
\label{alg: Heuristic}
\end{algorithm}

Based on the aforementioned discussion, we illustrate the algorithm in \texttt{Algorithm~\ref{alg: Heuristic}}. The main idea is to sequentially increase the bit-rate demand until the provisioning capacity cannot satisfy it.   
 Specifically, in line~\ref{alg: step}, we manually set the throughput increment $\Delta_{TH}$ \textcolor{black}{($\Delta_{TH}=$25~Gbps is used, initial network throughput $TH$=0~Gbps)}. From lines~\ref{alg: start} to \ref{alg: end}, we try to increase the provisioning capacity  to satisfy the bit-rate demand of node pair $(s,d)$ by sequentially using options i), ii), and iii). Specifically, in line~\ref{alg: option 1}, we use option i) to increase the provisioning traffic. Then, either option ii) or iii) is adopted if the provisioning capacity is smaller than the bit-rate demand, as shown from lines~\ref{alg: if1} to \ref{alg: option 3}.  Finally, the algorithm terminates if  the maximum provisioning capacity is smaller than 
the bit-rate demand, as shown in line~\ref{alg: terminate}. \textcolor{black}{It should be noted that the three different options on the lightpath are all implemented among the candidate lightpath set $\mathcal{P}$. Thus, the different network traffic loads can be implemented by restricting the available channel indexes $ch$.}

\section{Simulation Results}\label{sec: simulation}

In this section, we use numerical simulations to verify the throughput improvement by optimizing channel spacing and using just-enough SNR margin. 
\textcolor{black}{First, we compare the throughput difference between just-enough SNR margin and excessive SNR margin in single baud-rate networks. The comparison of channel spacing strategies \textit{FIX}, \textit{CAN(opt)}, \textit{CAN(random)}, and the proposed \textit{CSO} is also made in terms of throughput. Next, we investigate the throughput difference in flexible baud-rate networks leveraging just-enough SNR margin provisioning and CSO, where we highlight the throughput of different baud-rate selection strategies. Finally, we investigate the impact of power setting for different channel spacing strategies, and analyze the cost-performance ratio or the feedback tuning algorithm.}

\subsection{Simulation setup}\label{sec: simulation_setup}

The simulations use three network topologies, including $4$-node ring (400~km per link), Cost239 network (11 nodes, 52 links)\cite{JZZX16}, and NSF network (14 nodes, 44 links)\cite{JZZX16}. \textcolor{black}{It should be noted that the NSF network are scaled with a factor of 0.5 to guarantee 
the traffic of the longest shortest path can be transmitted.} To guarantee that the ILP models in \texttt{Phase 2} and \texttt{Phase 3} are solvable within a reasonable time, we assume that the available spectrum resource per link of 4-node ring is $F$=~750~GHz ($W$=~60~FSs). While in the mesh network, we assume the available spectrum resource is $F$=~3,000~GHz ($W$=~240~FSs). Other network parameters are stated as follows. The network traffic load is varied with the available spectrum resources, $W_{cur}$ from 20\%$\times W$ to 100\%$\times W$. The number of candidate routes $K$ equals 10. \textcolor{black}{Similar to the prior studies\cite{IvBS15,YADW17,ZhWA15}, we adopt the widely used assumption of uniform traffic for the normalized demand matrix}, i.e. $\hat{D}_{s,d}=\frac{1}{n\cdot(n-1)}$. 

Our simulations run on an Intel Core PC with 3.6~GHz CPU and 64~GB RAM. Specifically,  the LP/ILP models and heuristic algorithm are solved  with Gurobi 9.0 and MATLAB 2017a. For each LP/ILP model, we set the optimality gap of 5\% to guarantee the computational efficiency.  The typical single-mode fiber is assumed for physical layer calculation, i.e. fiber attenuation ratio $\alpha$=0.2~dB/km, second order dispersion coefficient $|\beta_2|$=21.7~ps$^2$/km, frequency of optical signal $\nu=192.5$~THz, and fiber nonlinear coefficient $\gamma$=1.3~(W$\cdot$Km)$^{-1}$. For the optical amplifiers, the noise figure $n_{sp}=5$~dB and $L_{span}$=100~km. \textcolor{black}{The PSD for all transceivers is assumed with a constant 25~$\mu$W/GHz, which is slightly larger than the power calculated by the LOGON strategy (15.3~$\mu$W/GHz among 3,000~GHz). The impact of different PSD settings  will be also discussed in  Sec.~\ref{sec: pli_disucss}.} In this paper, we have adopted four different MFs (PM-QPSK, PM-16QAM, and PM-64QAM, PM-256QAM) and at most five FEC overheads for each MF so that we can obtain the bit-rate from 50 to 375 Gbps with a step of 25 Gbps\cite{IWLS16}. Besides, the SNR threshold of each transmission mode has been individually superimposed with a \textcolor{black}{fixed penalty compared to the original data in \cite{IWLS16} in order to avoid the unexpected electrical noise (receiver and thermal shot noise, etc).} Three baud-rates are also assumed\cite[Table 3]{OIF2019}: 16, 32, and 64~Gbaud, \textcolor{black}{which occupies 2, 4, and 6 FSs, respectively. Although the SNR threshold of higher baud rates could be introduced due to the current hardware implementation\cite{GSSE20}, this paper assumes an ideal case that the SNR threshold for different baud-rates is identical.}
 
For ease of illustration, we use JP and EP to denote the just-enough SNR margin provisioning and traditional excessive SNR margin provisioning. \textcolor{black}{The statistics value of the channel spacing for each lightpath is defined as the \textit{minimum channel spacing with other lightpaths}.} 

\subsection{JP vs. traditional EP in single baud-rate networks}
First, we compare the throughput difference between JP and EP \textcolor{black}{for networks adopting a single baud-rate (16 Gbaud). The latter is implemented by breaking the iterative loop of Fig.~\ref{fig: flowchart}.} \textcolor{black}{The results from different lightpath provisioning algorithms are compared in order to validate the performance of JP, including the heuristic in \texttt{Algorithm~\ref{alg: Heuristic}}, and the ILP models in \texttt{Phase 2} and \texttt{Phase 3}.} 
\subsubsection{JP(CSO) vs.  traditional EP(CSO) in a small network}


\begin{figure}[!htbp]
    \centering
    \includegraphics[]{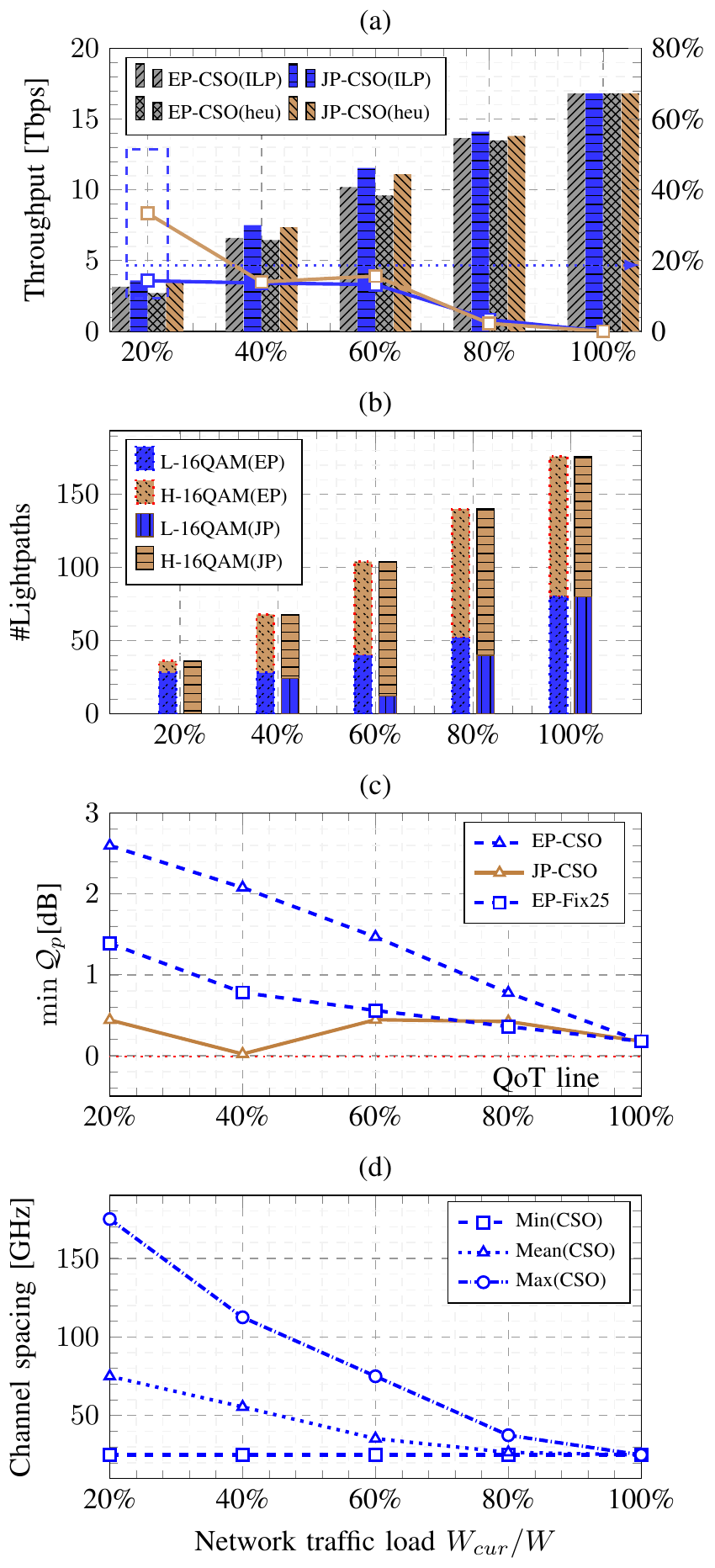}
    \caption{Comparison of  JP and EP with CSO for 4-node ring. (a) Throughput vs. network traffic load. The relative gain ratio between JP and EP is also plotted on the right $y$-axis. (b) Adopted transmission modes vs. network traffic load. L-16QAM includes 16QAM with FEC 0.62 and 0.72, while H-16QAM includes 16QAM with FEC 0.82 and 0.92. The spectral efficiency of H-16QAM is higher than L-16QAM; (c) Minimum QoT metric vs. network traffic load;  (d) Channel spacing statistics of CSO.}
    \label{fig: channel_spacing_ring}
\end{figure}

\textcolor{black}{
Figure~\ref{fig: channel_spacing_ring} illustrates the throughput, the number of adopted transmission modes, the QoT metric, and the lightpath's channel spacing when network traffic load increases. Regardless of the different lightpath provisioning algorithms,  JP outperforms the EP under the resource over-provisioning scenarios in terms of the throughput (network traffic load between 20\% and 80\% in Fig.~\ref{fig: channel_spacing_ring}(a)). The observed increase in throughput is attributed to the more usage of high-order transmission modes with high spectral efficiency (H-16QAM) (Fig.~\ref{fig: channel_spacing_ring}(b)). 
}

\textcolor{black}{
Another observation in Fig.~\ref{fig: channel_spacing_ring}(a) is that the relative throughput gain of JP decreases with the network traffic load. This trend  could be explained using the conclusion of \cite[Sec.~5 and 6]{SaVI19} that the fraction ratio of achievable capacity is approximately linearly with the overestimation of network SNR margin.  As network traffic load increases, less SNR margin overestimation (Fig.~\ref{fig: channel_spacing_ring}(c))  between resource over-provisioning and resource under-provisioning  means less throughput gain (Fig.~\ref{fig: channel_spacing_ring}(a)).
It should also be mentioned that the higher throughput of JP is achieved by using the high spectral efficiency transmission mode without increasing the number of lightpaths (Fig.~\ref{fig: channel_spacing_ring}(b)). However, it cannot be ruled out that in some special cases, the heuristic algorithm may generate extra lightpaths even under the same network traffic load to obtain the higher throughput. We will analyze the cost-performance ratio in Sec.~\ref{sec: cost-performance}.}

\textcolor{black}{
In Fig.~\ref{fig: channel_spacing_ring}(d), we find that the channel spacing statistics value (minimum channel spacing, average channel spacing, and maximum channel spacing) varies with the different network traffic loads. Such results imply that the traditional fixed channel spacing probably lacks robustness when setting the channel spacing. Next, we will discuss the resulting network performance difference of this characteristic.
}

\subsubsection{JP(FIX, CAN, CSO) vs. traditional EP(CSO) in large networks}\label{sec: JP EP in large}

\begin{figure*}[!htbp]
\centering 
\subfigure[Throughput (NSF)]{
\includegraphics[height=4.5cm]{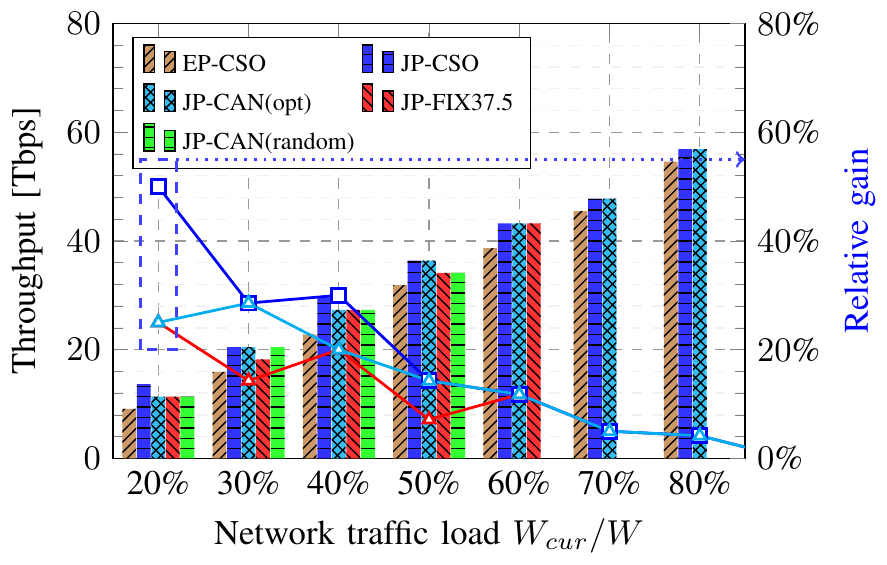}
}
\subfigure[Channel spacing under 20\% load (NSF) ]{
\includegraphics[width=4cm]{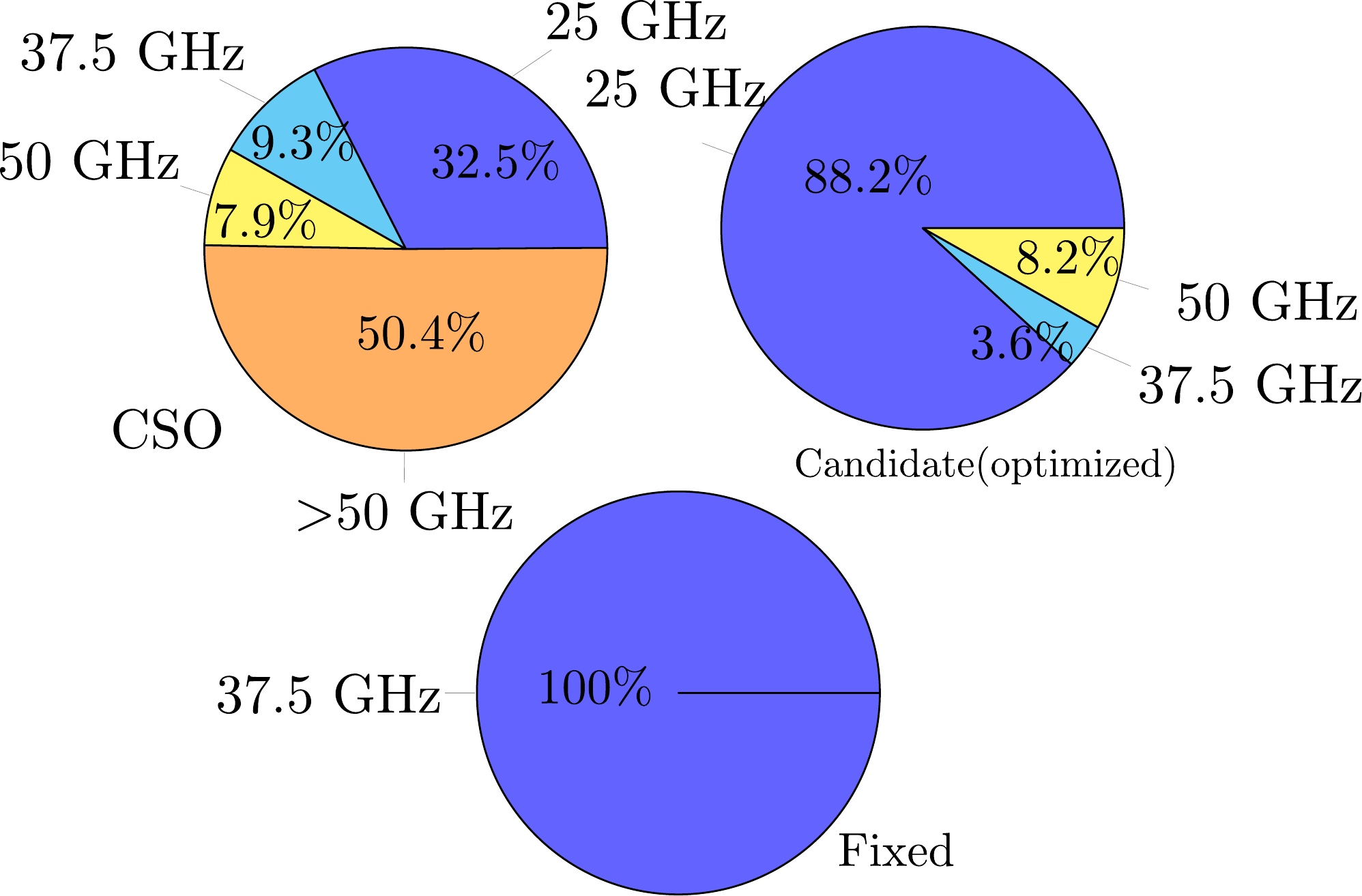}
} 
\subfigure[Link spectrum usage (20\% load)  (NSF)]{
\includegraphics[width=5.8cm]{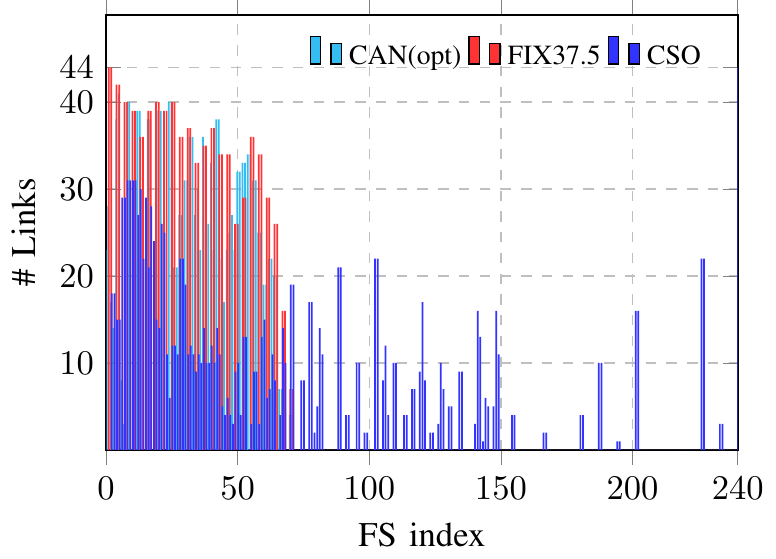}
}
\subfigure[Throughput (Cost239)]{
\includegraphics[height=4.5cm]{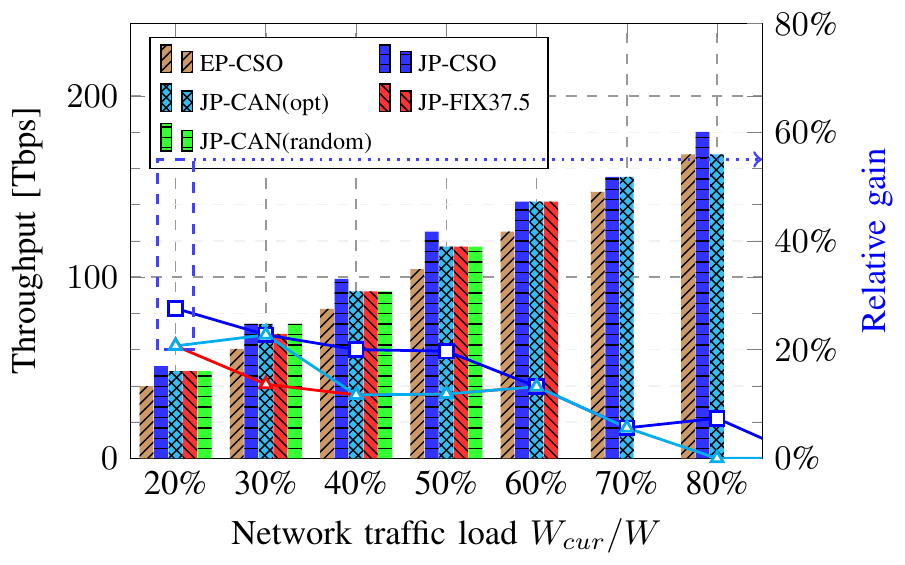}
}
\subfigure[Channel spacing under 20\% load (Cost239) ]{
\includegraphics[width=4cm]{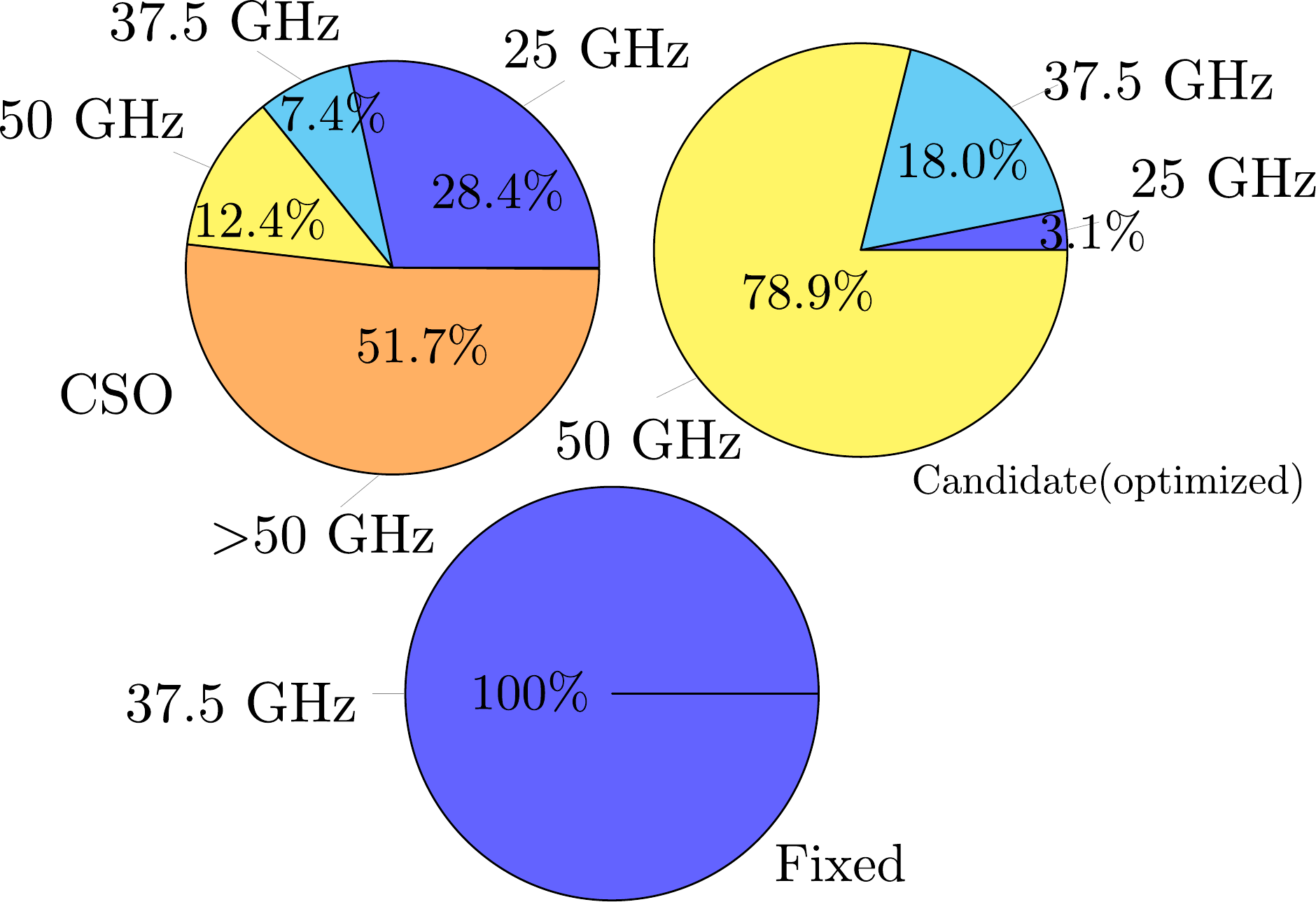}
} 
\subfigure[Link spectrum usage (20\% load) (Cost239)]{
\includegraphics[width=5.8cm]{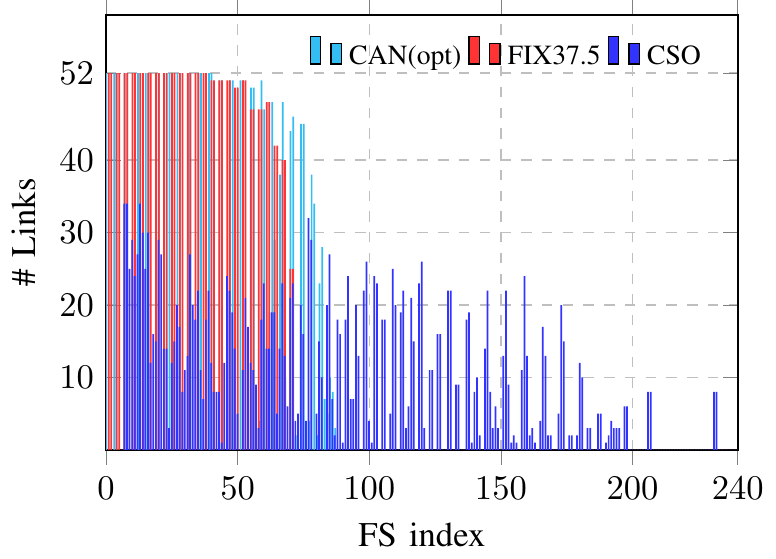}
}
    \caption{Throughput with different channel spacing strategies.}
    \label{fig: channel_spacing_optimziation}
\end{figure*}

\textcolor{black}{
Next, we investigate the throughput of different channel spacing strategies. $\mathcal{H}=\{37.5\}~\text{GHz}$ for FIX, $\mathcal{H}=\{25, 37.5, 50\}~\text{GHz}$ for CAN(opt) and CAN(random).
}

\textcolor{black}{
In Fig.~\ref{fig: channel_spacing_optimziation}(a), 
the largest throughput gain ratio (50\%)  at the network traffic load of 20\% is obtained by CSO among the existing channel spacing strategies (FIX, CAN, CSO). This is expected because CSO uses no artificial constraint of channel spacing and allows the transceiver to use the idle spectrum resources to enhance the physical layer performance  and upgrade the high-order transmission mode. To gain more insight, we plot the number of adopted channel spacing  in Fig.~\ref{fig: channel_spacing_optimziation}(b) as well as the link statistics on each FS in Fig.~\ref{fig: channel_spacing_optimziation}(c). We see that the most significant channel spacing ($\geq$50~GHz) is out of the scope of the other strategies (FIX and CAN). The above findings can also be  observed in the Cost239 network in Figs.~\ref{fig: channel_spacing_optimziation}(e). 
}

\subsection{JP vs. traditional EP in multiple baud-rates networks}

\textcolor{black}{
Next, we study the throughput of different baud-rates and compare the baud-rate selection policies (Fig.~\ref{fig: BVT_Cost239}).  The comparison of different channel spacing strategies is also made (Table~\ref{tab: CSO_in}).
}

\subsubsection{Different single baud-rates}

\textcolor{black}{The network throughput of three different single baud-rates is shown in Fig.~\ref{fig: BVT_Cost239}(a).  At the network traffic load of 100\%, the largest throughput (295 Tbps) is achieved by the 64 Gbaud among three baud-rates (210~Tbps for 16 and 32 Gbaud).
It is expected because the spectral efficiency of 64 Gbaud transceiver is 2.7 bps/Hz, while it is 2.0 bps/Hz for both 16 and 32 Gbaud (50~Gbps for 16 Gbaud on 25~GHz, 100~Gbps for 32 Gbaud on 50~GHz, 200~Gbps for 64 Gbaud on 75~GHz \cite{IWLS16, OIF2019}). 
An interesting observation in Fig.~\ref{fig: BVT_Cost239}(b) is that the maximal absolute gain (33/22/19.5 Tbps) is achieved in the medium network traffic load for these single baud-rate networks. We provide a possible explanation by incorporating the previous finding on the throughput gain ratio in Figs.~\ref{fig: channel_spacing_ring} and \ref{fig: channel_spacing_optimziation}. 
On the one hand, the available spectrum resources
increase with the network traffic load. On the other hand, the relative throughput gain ratio  (or relative net spectral efficiency) decreases (see Figs.~\ref{fig: channel_spacing_optimziation}(a) and (d)). Therefore, their product, namely absolute throughput gain, could achieve the maximum when the network traffic load is at a certain value (medium in this case).
}

\subsubsection{Baud-rate selection strategies (low baud-rate first, high baud-rate first, vs. random baud-rate)}
\begin{figure*}[!htbp]
    \centering
\includegraphics[width=0.98\textwidth]{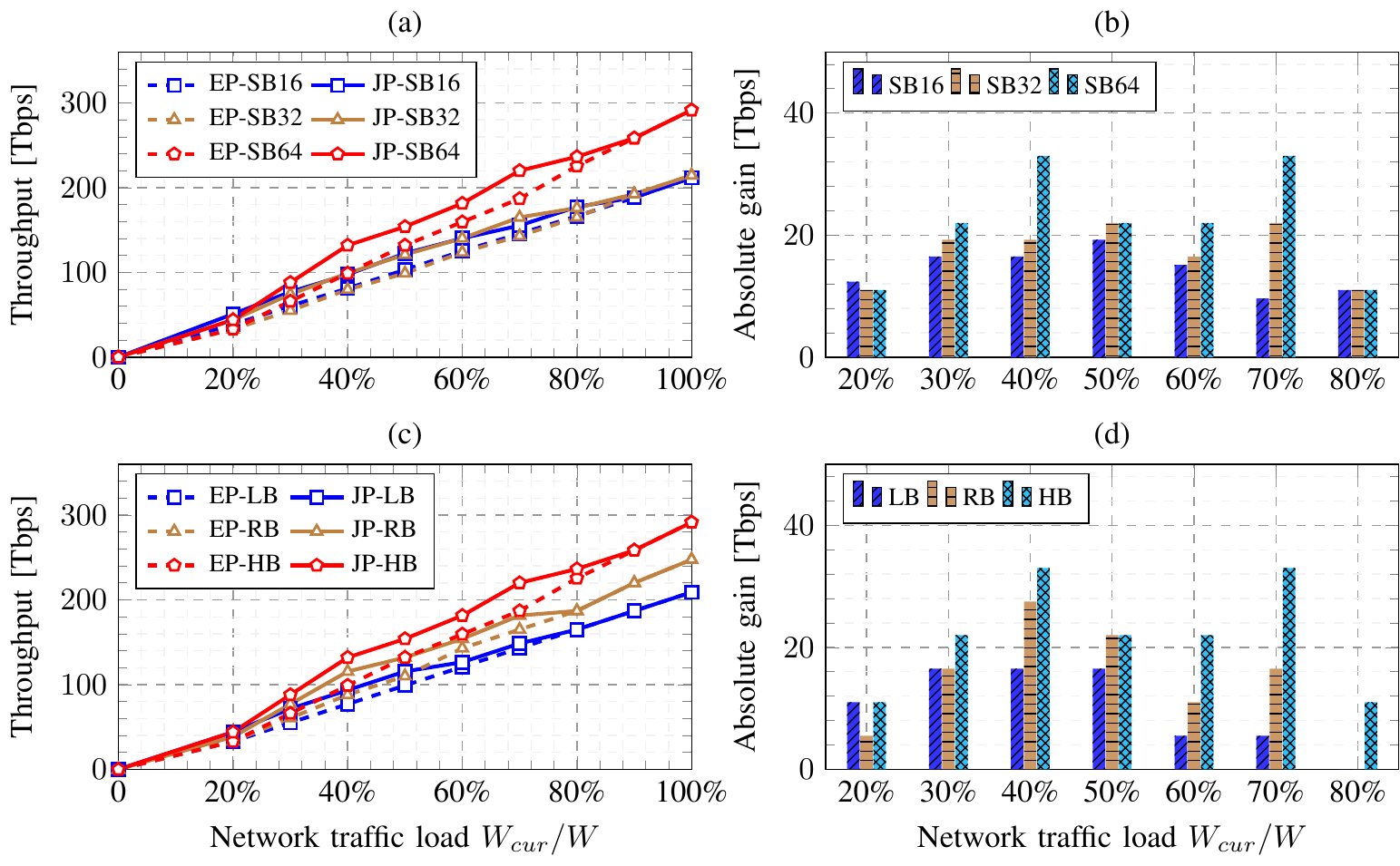}  
    \caption{Throughput in Cost239 network with different baud-rate configurations. (a) Different single baud-rates (SB16: only single 16 Gbaud); (b) Absolute throughput gain of JP with single baud-rate; (c) Different selection strategies of baud-rates (LB: low baud-rate first. RB: random baud-rate. HB: high baud-rate first); (d) Absolute throughput gain of JP with different  selection strategies.}
    \label{fig: BVT_Cost239}
\end{figure*}

\textcolor{black}{Figure~\ref{fig: BVT_Cost239}(c) shows the network throughput using different baud-rate selection policies. The utilization of prioritizing the higher baud-rate (higher baud-rate means higher spectral efficiency) has a higher throughput, which is similar to the study \cite{PCSP19} that priorities the maximum spectral efficient channels. Such a result validates the efficient selection strategy to achieve the higher throughput from another perspective.
}

\subsubsection{JP(FIX, CAN, CSO) vs. traditional EP(CSO)}
\textcolor{black}{ We compare the throughput gain ratio of different channel spacing strategies in networks supporting flexible baud-rates. Table~\ref{tab: CSO_in} illustrates the relative throughput gain ratio of different channel spacing strategies. The performance of CAN (opt) is closer to the proposed CSO if the size of candidate channel spacing set increases. However, the FIX and CAN (random) can be applied to increase the network throughput only in the scenarios of low network traffic load. These results show the advantage of CSO with a larger throughput gain ratio and with more network traffic load scenarios. 
}

\begin{table}[!htbp]   \caption{Relative throughput gain ratio in Cost239 network with single baud-rate (above) and with flexible baud-rates (below)}
\begin{threeparttable}
\centering
  \includegraphics[width=0.5\textwidth]{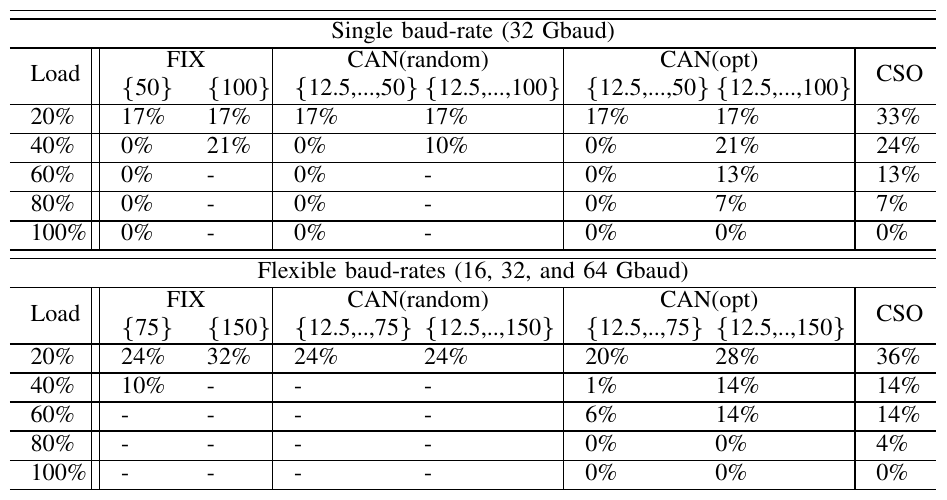}
    \label{tab: CSO_in}
    \scriptsize
\begin{tablenotes}
\item[-] It signifies the channel spacing strategy is infeasible due to the limited spectrum resources.
\end{tablenotes}
\end{threeparttable}
\end{table}

\subsection{Cost-performance ratio}\label{sec: cost-performance}

\begin{figure}[!htbp]
    \centering
    \includegraphics[]{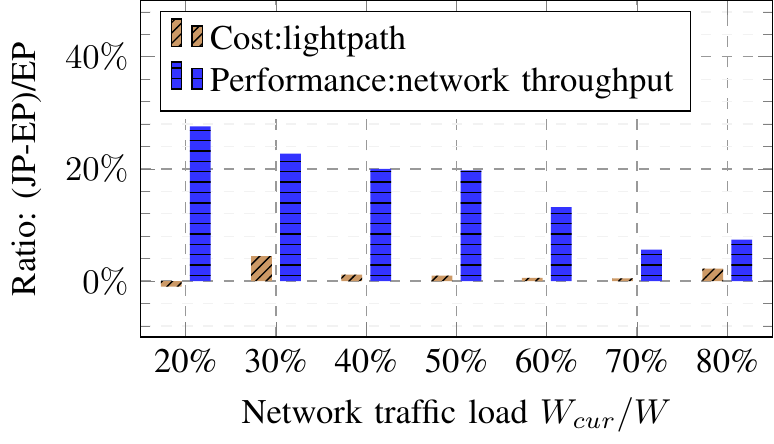}
    \caption{Cost-performance ratio of JP-CSO and EP in Cost239 network.}
    \label{fig: cost}
\end{figure}

\textcolor{black}{
In addition, we analyze the cost-performance ratio  of the feedback tuning algorithm. 
Taken the Cost239 network as an example, the relative increase of lightpath (all adopt 16~Gbaud) and throughput from EP to JP are shown in Fig.~\ref{fig: cost}. The throughput gain ratio varies from 7\% to 27\%, which is always higher than the cost ratio that varies from -1\% to 5\%. Such results reveal that the JP is more cost-effective.
}

\subsection{Impact of physical layer parameters}\label{sec: pli_disucss}
 
\begin{figure}[!htbp]
    \centering
    \includegraphics{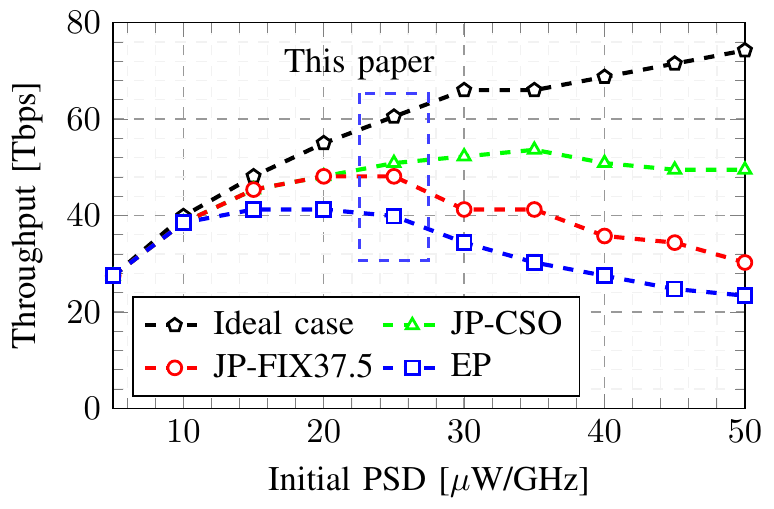}
    \caption{Throughput vs. different initial PSDs under 20\% network loads in Cost239 network (PSD=25$\mu$W/GHz is adopted in the simulations). ``Ideal case'' means no NLI in the networks.}
    \label{fig: Impact of PSD}
\end{figure}
 
\textcolor{black}{
This section discusses the impact of different initial PSDs on channel spacing optimization.  In Fig.~\ref{fig: Impact of PSD}, we see that the throughput increases with the initial PSD in the linear regime where the PSD is lower than 15~$\mu$W/GHz, and decreases in the nonlinear regime where the PSD exceeds 15~$\mu$W/GHz. The higher throughput is maintained if we adopt the CSO, especially in the nonlinear regime. Such results imply that the CSO can tolerate a higher PSD where the NLI is more significant (for example, the case of network being attacked\cite{KFZW16}). 
}

\section{Conclusion}\label{sec: conclusions}

In this paper, we studied the problem of \textit{throughput maximization leveraging just-enough SNR margin provisioning and channel spacing optimization}. Based on the analysis of the lightpath's parameters, we developed an iterative feedback tuning algorithm to provide a just-enough SNR margin. We also established an ILP model and a heuristic algorithm to maximize the network throughput \textcolor{black}{by optimizing  route, MF, FEC, baud rate, and spectrum assignment}. Furthermore, we devised  \textcolor{black}{a low-complexity} LP model that enables us to optimize the channel spacing for \textcolor{black}{a large number of lightpaths}.

Through the simulations, we verify the throughput improvement of just-enough SNR margin provisioning compared to the conventional excessive SNR margin provisioning. The relative gain reaches the maximum (over 20\%) under low network traffic load, while the absolute gain reaches the maximum under medium network traffic load. These observations highlight that the importance of just-enough SNR margin provisioning under low and medium network traffic load. Compared to the existing channel spacing strategies (fixed and candidate), our channel spacing optimization also demonstrates a clear advantage on two aspects: (1) it can be applied with more resource over-provisioning scenarios; (2) it has a larger relative throughput gain ratio. \textcolor{black}{We also show that the channel spacing becomes more significant when the NLI from neighbor channels becomes severe, i.e. in the nonlinear regime. In addition, we} find that the higher baud-rate with larger spectral efficiency should be prioritized to gain a larger throughput when implementing the lightpath provisioning. Taken together, our findings highlight an important role for just-enough SNR margin provisioning and channel spacing optimization in FONs with flexible baud-rates.

\bibliographystyle{IEEEtran}
\bibliography{reference20200507}

\begin{thebibliography}{10}
\providecommand{\url}[1]{#1}
\csname url@samestyle\endcsname
\providecommand{\newblock}{\relax}
\providecommand{\bibinfo}[2]{#2}
\providecommand{\BIBentrySTDinterwordspacing}{\spaceskip=0pt\relax}
\providecommand{\BIBentryALTinterwordstretchfactor}{4}
\providecommand{\BIBentryALTinterwordspacing}{\spaceskip=\fontdimen2\font plus
\BIBentryALTinterwordstretchfactor\fontdimen3\font minus
  \fontdimen4\font\relax}
\providecommand{\BIBforeignlanguage}[2]{{%
\expandafter\ifx\csname l@#1\endcsname\relax
\typeout{** WARNING: IEEEtran.bst: No hyphenation pattern has been}%
\typeout{** loaded for the language `#1'. Using the pattern for}%
\typeout{** the default language instead.}%
\else
\language=\csname l@#1\endcsname
\fi
#2}}
\providecommand{\BIBdecl}{\relax}
\BIBdecl

\bibitem{cisco2017}
``Cisco visual networking index: Forecast and trends, 2017-2022,'' Cisco, Tech.
  Rep., Nov. 2018.

\bibitem{SGSC20}
A.~Sgambelluri, A.~Giorgetti, D.~Scano, F.~Cugini, and F.~Paolucci,
  ``Openconfig and openroadm automation of operational modes in disaggregated
  optical networks,'' \emph{IEEE Access}, vol.~8, pp. 190\,094--190\,107, Oct.
  2020.

\bibitem{Bosc19}
G.~{Bosco}, ``Advanced modulation techniques for flexible optical transceivers:
  The rate/reach trade-off,'' \emph{IEEE/OSA J. Lightw. Technol.}, vol.~37,
  no.~1, pp. 36--49, Jan. 2019.

\bibitem{DaLS15}
H.~Dai, Y.~Li, and G.~Shen, ``Explore maximal potential capacity of {WDM}
  optical networks using time domain hybrid modulation technique,''
  \emph{IEEE/OSA J. Lightwave Technol.}, vol.~33, no.~18, pp. 3815--3826, May
  2015.

\bibitem{IvBS15}
D.~J. Ives, P.~Bayvel, and S.~J. Savory, ``Routing, modulation, spectrum and
  launch power assignment to maximize the traffic throughput of a nonlinear
  optical mesh network,'' \emph{Photon. Netw. Commun.}, vol.~29, no.~3, pp.
  244--256, Jun. 2015.

\bibitem{BeSa13}
H.~Beyranvand and J.~A. Salehi, ``A quality-of-transmission aware dynamic
  routing and spectrum assignment scheme for future elastic optical networks,''
  \emph{IEEE/OSA J. Lightw. Technol.}, vol.~31, no.~18, pp. 3043--3054, 2013.

\bibitem{XYAP19}
Y.~{Xu}, L.~{Yan}, E.~{Agrell}, and M.~{Brandt-Pearce}, ``Iterative resource
  allocation algorithm for {EONs} based on a linearized {GN} model,''
  \emph{IEEE/OSA J. Opt. Commun. Netw.}, vol.~11, no.~3, pp. 39--51, Mar. 2019.

\bibitem{YADW17}
L.~Yan, E.~Agrell, M.~N. Dharmaweera, and H.~Wymeersch, ``Joint assignment of
  power, routing, and spectrum in static flexible-grid networks,''
  \emph{IEEE/OSA J. Lightw. Technol.}, vol.~35, no.~10, pp. 1766--1774, May
  2017.

\bibitem{WBMN19}
R.~{Wang}, S.~{Bidkar}, F.~{Meng}, R.~{Nejabati}, and D.~{Simeonidou},
  ``Load-aware nonlinearity estimation for elastic optical network resource
  optimization and management,'' \emph{IEEE/OSA J. Opt. Commun. Netw.},
  vol.~11, no.~5, pp. 164--178, May 2019.

\bibitem{ZhWA15}
J.~Zhao, H.~Wymeersch, and E.~Agrell, ``Nonlinear impairment-aware static
  resource allocation in elastic optical networks,'' \emph{IEEE/OSA J. Lightw.
  Technol.}, vol.~33, no.~22, pp. 4554--4564, Aug. 2015.

\bibitem{Yvan17}
Y.~Pointurier, ``Design of low-margin optical networks,'' \emph{IEEE/OSA J.
  Opt. Commun. Netw.}, vol.~9, no.~1, pp. A9--A17, Jan. 2017.

\bibitem{SaVI19}
S.~J. Savory, R.~J. Vincent, and D.~J. Ives, ``Design considerations for
  low-margin elastic optical networks in the nonlinear regime [invited],''
  \emph{IEEE/OSA J. Opt. Commun. Netw.}, vol.~11, no.~10, pp. C76--C85, Oct.
  2019.

\bibitem{Ciena19}
\BIBentryALTinterwordspacing
Ciena, \emph{Transforming Margin into Capacity with Liquid Spectrum}, (Date
  last accessed 2-February-2021). [Online]. Available:
  \url{https://media.ciena.com/documents/Transforming_Margin_into_Capacity_with_Liquid_Spectrum_WP.pdf}
\BIBentrySTDinterwordspacing

\bibitem{Infinera19}
\BIBentryALTinterwordspacing
Infinera, \emph{The ultimate guide to higher baud rates.}, (Date last accessed
  2-November-2020). [Online]. Available:
  \url{https://www.infinera.com/white-paper/The-Ultimate-Guide-to-Higher-Baud-Rates}
\BIBentrySTDinterwordspacing

\bibitem{MRZG15}
A.~{Morea}, J.~{Renaudier}, T.~{Zami}, A.~{Ghazisaeidi}, and
  O.~{Bertran-Pardo}, ``Throughput comparison between 50-{GHz} and 37.5-{GHz}
  grid transparent networks,'' \emph{IEEE/OSA J. Opt. Commun. Netw.}, vol.~7,
  no.~2, pp. A293--A300, Feb. 2015.

\bibitem{CMLB17}
J.~C. Cartledge, F.~Matos, C.~Laperle, A.~Borowiec, M.~O'Sullivan, and
  K.~Roberts, ``Use of extreme value statistics to assess the performance
  implications of cascaded {ROADMs},'' \emph{IEEE/OSA J. Lightw. Technol.},
  vol.~35, no.~23, pp. 5208--5214, Dec. 2017.

\bibitem{GSSE20}
T.~Gerard, D.~Semrau, E.~Sillekens, A.~Edwards, W.~Pelouch, S.~Barnes, R.~I.
  Killey, D.~Lavery, P.~Bayvel, and L.~Galdino, ``Relative impact of channel
  symbol rate on transmission capacity,'' \emph{IEEE/OSA J. Opt. Commun.
  Netw.}, vol.~12, no.~4, pp. B1--B8, Jan. 2020.

\bibitem{RBMT17}
C.~Rottondi, P.~Boffi, P.~Martelli, and M.~Tornatore, ``Routing, modulation
  format, baud rate and spectrum allocation in optical metro rings with
  flexible grid and few-mode transmission,'' \emph{IEEE/OSA J. Lightw.
  Technol.}, vol.~35, no.~1, pp. 61--70, Jan. 2017.

\bibitem{STCT19}
N.~{Shahriar}, S.~{Taeb}, S.~R. {Chowdhury}, M.~{Tornatore}, R.~{Boutaba},
  J.~{Mitra}, and M.~{Hemmati}, ``Achieving a fully-flexible virtual network
  embedding in elastic optical networks,'' in \emph{Proc. Int. Conf. on Comput.
  Commun. (ICC)}, Paris, France, Apr. 2019, pp. 1756--1764.

\bibitem{SMCF15}
N.~Sambo, G.~Meloni, F.~Cugini, F.~Fresi, A.~D'Errico, L.~Poti, P.~Iovanna, and
  P.~Castoldi, ``Routing, code, and spectrum assignment, subcarrier spacing,
  and filter configuration in elastic optical networks,'' \emph{IEEE/OSA J.
  Opt. Commun. Netw.}, vol.~7, no.~11, pp. B93--B100, Nov. 2015.

\bibitem{PCSP19}
J.~{Pedro}, N.~{Costa}, B.~{Sommerkorn-Krombholz}, and S.~{Pato}, ``Capacity
  limits of mesh optical transport networks exploiting future high baud-rate
  line interfaces,'' in \emph{Proc. Int. Conf. Transparent Opt. Netw.}, Angers,
  France, Jul. 2019, pp. 1--7.

\bibitem{SaCV18}
I.~{Sartzetakis}, K.~{Christodoulopoulos}, and E.~{Varvarigos}, ``Cross-layer
  adaptive elastic optical networks,'' \emph{IEEE/OSA J. Opt. Commun. Netw.},
  vol.~10, no.~2, pp. A154--A164, Feb. 2018.

\bibitem{SCQP17}
P.~{Soumplis}, K.~{Christodoulopoulos}, M.~{Quagliotti}, A.~{Pagano}, and
  E.~{Varvarigos}, ``Network planning with actual margins,'' \emph{IEEE/OSA J.
  Lightw. Technol.}, vol.~35, no.~23, pp. 5105--5120, Oct. 2017.

\bibitem{Pedr17}
J.~{Pedro}, ``Designing transparent flexible-grid optical networks for maximum
  spectral efficiency [invited],'' \emph{IEEE/OSA J. Opt. Commun. Netw.},
  vol.~9, no.~4, pp. C35--C44, Apr. 2017.

\bibitem{Auge13}
J.~{Aug\'e}, ``Can we use flexible transponders to reduce margins?'' in
  \emph{Proc. Opt. Fiber Commun. Conf. (OFC)}, Anaheim, USA, Mar. 2013, p.
  OTu2A.1.

\bibitem{PBCK13}
E.~Palkopoulou, G.~Bosco, A.~Carena, D.~Klonidis, P.~Poggiolini, and I.~Tomkos,
  ``Nyquist-{WDM}-based flexible optical networks: Exploring physical layer
  design parameters,'' \emph{IEEE/OSA J. Lightwave Technol.}, vol.~31, no.~14,
  pp. 2332--2339, Jul. 2013.

\bibitem{Pogg12}
P.~Poggiolini, ``The {GN} model of non-linear propagation in uncompensated
  coherent optical systems,'' \emph{IEEE/OSA J. Lightw. Technol.}, vol.~30,
  no.~24, pp. 3857--3879, Dec. 2012.

\bibitem{JoAg14}
P.~Johannisson and E.~Agrell, ``Modeling of nonlinear signal distortion in
  fiber-optic networks,'' \emph{IEEE/OSA J. Lightw. Technol.}, vol.~32, no.~23,
  pp. 3942--3950, Oct. 2014.

\bibitem{RoKB16}
I.~Roberts, J.~M. Kahn, and D.~Boertjes, ``Convex channel power optimization in
  nonlinear {WDM} systems using {Gaussian} {Noise} model,'' \emph{IEEE/OSA J.
  Lightw. Technol.}, vol.~34, no.~13, pp. 3212--3222, May 2016.

\bibitem{BAKK19}
C.~{Bhar}, E.~{Agrell}, K.~{Keykhosravi}, M.~{Karlsson}, and P.~A. {Andrekson},
  ``Channel allocation in elastic optical networks using traveling salesman
  problem algorithms,'' \emph{IEEE/OSA J. Opt. Commun. Netw.}, vol.~11, no.~10,
  pp. C58--C66, Oct. 2019.

\bibitem{YAWP15}
L.~Yan, E.~Agrell, H.~Wymeersch, and M.~Brandt-Pearce, ``Resource allocation
  for flexible-grid optical networks with nonlinear channel model,''
  \emph{IEEE/OSA J. Opt. Commun. Netw.}, vol.~7, no.~11, pp. B101--B108, Nov.
  2015.

\bibitem{RRBZ20}
A.~Raeesi, H.~Rabbani, L.~Beygi, and S.~Zokaei, ``Low-complexity physical layer
  impairment aware spectrum assignment based on discretized gaussian model for
  nonlinear noise in elastic optical networks,'' \emph{Optics Commun.}, vol.
  474, p. 126011, Apr. 2020.

\bibitem{KFZW16}
N.~Skorin-Kapov, M.~Furdek, S.~Zsigmond, and L.~Wosinska, ``Physical-layer
  security in evolving optical networks,'' \emph{IEEE Commun. Mag.}, vol.~54,
  no.~8, pp. 110--117, Aug. 2016.

\bibitem{BCCP11}
G.~Bosco, V.~Curri, A.~Carena, P.~Poggiolini, and F.~Forghieri, ``On the
  performance of {Nyquist-WDM} terabit superchannels based on {PM-BPSK,
  PM-QPSK, PM-8QAM or PM-16QAM} subcarriers,'' \emph{IEEE/OSA J. Lightw.
  Technol.}, vol.~29, no.~1, pp. 53--61, Jan. 2011.

\bibitem{NaTM13}
A.~Nag, M.~Tornatore, and B.~Mukherjee, ``On the effect of channel spacing,
  launch power, and regenerator placement on the design of mixed-line-rate
  optical networks,'' \emph{Opt. Switch and Netw.}, vol.~10, no.~4, pp. 301 --
  311, Nov. 2013.

\bibitem{ONDM2020}
C.~Chen, F.~Zhou, Y.~Liu, and S.~Xiao, ``Channel frequency optimization in
  optical networks based on {Gaussian Noise} model,'' in \emph{Proc. Int. Conf.
  Opt. Netw. Design and Modeling (ONDM)}, Barcelona, Spain, May 2020, pp. 1--6.

\bibitem{OIF2019}
\BIBentryALTinterwordspacing
J.~D. Reis, V.~Shukla, D.~R. Stauffer, and K.~Gass, ``Technology options for
  {400G} implementation,'' The Optical Internetworking Forum, Tech. Rep., Jul.
  2015. [Online]. Available:
  \url{https://www.oiforum.com/wp-content/uploads/2019/01/OIF-Tech-Options-400G-01.0.pdf}
\BIBentrySTDinterwordspacing

\bibitem{IWLS16}
D.~J. {Ives}, P.~{Wright}, A.~{Lord}, and S.~J. {Savory}, ``Using 25 gbe client
  rates to access the gains of adaptive bit- and code-rate networking,''
  \emph{IEEE/OSA J. Opt. Commun. Netw.}, vol.~8, no.~7, pp. A86--A91, Jul.
  2016.

\bibitem{IvBS14}
D.~J. {Ives}, P.~{Bayvel}, and S.~J. {Savory}, ``Adapting transmitter power and
  modulation format to improve optical network performance utilizing the
  {Gaussian Noise} model of nonlinear impairments,'' \emph{IEEE/OSA J. Lightw.
  Technol.}, vol.~32, no.~21, pp. 4087--4096, Nov. 2014.

\bibitem{IpKa08}
E.~{Ip} and J.~M. {Kahn}, ``Compensation of dispersion and nonlinear
  impairments using digital backpropagation,'' \emph{IEEE/OSA J. Lightw.
  Technol.}, vol.~26, no.~20, pp. 3416--3425, Oct. 2008.

\bibitem{Yen71}
J.~Y. Yen, ``Finding the k shortest loopless paths in a network,'' \emph{Manag.
  Sci.}, vol.~17, no.~11, pp. 712--716, Jul. 1971.

\bibitem{VKRC12}
L.~Velasco, M.~Klinkowski, M.~Ruiz, and J.~Comellas, ``Modeling the routing and
  spectrum allocation problem for flexgrid optical networks,'' \emph{Photon.
  Netw. Comm.}, vol.~24, no.~3, pp. 177--186, Apr. 2012.

\bibitem{PeCP20}
J.~{Pedro}, N.~{Costa}, and S.~{Pato}, ``Optical transport network design
  beyond 100 {Gbaud} [invited],'' \emph{IEEE/OSA J. Opt. Commun. Netw.},
  vol.~12, no.~2, pp. A123--A134, Nov. 2020.

\bibitem{RBGR19}
H.~{Rabbani}, L.~{Beygi}, S.~{Ghoshooni}, H.~{Rabbani}, and E.~{Agrell},
  ``Quality of transmission aware optical networking using enhanced gaussian
  noise model,'' \emph{IEEE/OSA J. Lightw. Technol.}, vol.~37, no.~3, pp.
  831--838, Feb. 2019.

\bibitem{PBCC13}
P.~{Poggiolini}, G.~{Bosco}, A.~{Carena}, R.~{Cigliutti}, V.~{Curri},
  F.~{Forghieri}, R.~{Pastorelli}, and S.~{Piciaccia}, ``The {LOGON} strategy
  for low-complexity control plane implementation in new-generation flexible
  networks,'' in \emph{Proc. Opt. Fiber Commun. Conf. (OFC)}, Anaheim, USA,
  Mar. 2013, p. OW1H.3.

\bibitem{MaBo09}
A.~Magnani and S.~P. Boyd, ``Convex piecewise-linear fitting,''
  \emph{Optimization and Engineering}, vol.~10, no.~1, pp. 1--17, Mar. 2009.

\bibitem{JZZX16}
M.~{Ju}, F.~{Zhou}, Z.~{Zhu}, and S.~{Xiao}, ``Distance-adaptive, low {CAPEX}
  cost $p$-cycle design without candidate cycle enumeration in mixed-line-rate
  optical networks,'' \emph{IEEE/OSA J. Lightw. Technol.}, vol.~34, no.~11, pp.
  2663--2676, Apr. 2016.

\end{thebibliography}

\end{document}